\documentclass[11pt, a4paper]{article}
\usepackage{amsmath}
\usepackage{setspace}
\usepackage{bm}
\usepackage{graphicx}
\usepackage{subcaption}
\usepackage{bibentry}
\graphicspath{{Figures/}}
\usepackage{float}
\usepackage{hyperref}
\usepackage[round, authoryear, sort]{natbib}
\usepackage{mathptmx}
\usepackage{lmodern}
\usepackage{adjustbox}
\usepackage{rotating}
\usepackage[titletoc]{appendix}
\title{A Robust Bayesian Exponentially Tilted\\ Empirical Likelihood Method\thanks{The authors would like to thank 
		Eric Renault and David T. Frazier for very thoughtful comments
		during the development of this paper. We also thank participants at the Approximate Bayesian Computation and its Applications Workshop (ABC@ACEMS 2017), the 1st International Conference on Econometrics and Statistics Conference (EcoSta 2017), the 11th Conference on Bayesian NonParametrics (BNP11), the European Seminar on Bayesian Econometrics (ESOBE 2017), the 2017 International Workshop on Objective Bayes Methodology (OBayes 2017), the 11th International Conference on Computational and Financial Econometrics (CFE 2017) and seminars held at The Ohio State University, Brown University and Monash University. We gratefully acknowledge support provided by Australian Research Council Discovery Grant DP150101728.}}
\author{Zhichao Liu, Catherine S. Forbes and Heather M. Anderson}
\date{\today}
\begin{document}
	\maketitle
	\begin{abstract}
		This paper proposes a new Bayesian approach for analysing moment condition models in the situation where the data may be contaminated by outliers. The approach builds upon the foundations developed by \cite{Schennach-2005} who proposed the Bayesian exponentially tilted empirical likelihood (BETEL) method, justified by the fact that an empirical likelihood (EL) can be interpreted as the nonparametric limit of a Bayesian procedure when the implied probabilities are obtained from maximizing entropy subject to some given moment constraints. Considering the impact that outliers are thought to have on the estimation of population moments, we develop a new robust BETEL (RBETEL) inferential methodology to deal with this potential problem. We show how the BETEL methods are linked to the recent work of \cite{BHW-2016} who propose a general framework to update prior belief via a loss function. A controlled simulation experiment is conducted to investigate the performance of the RBETEL method. We find that the proposed methodology produces reliable posterior inference for the fundamental relationships that are embedded in the majority of the data, even when outliers are present. The method is also illustrated in an empirical study relating brain weight to body weight using a dataset containing sixty-five different land animal species.\\ 

	\end{abstract}
	\clearpage
\section{Introduction} \label{s:Intro}
Traditional parametric Bayesian analysis of data requires the specification of a likelihood function and a prior distribution for the parameters. However, when the data generating process (DGP) is unknown, meaning that the true likelihood function is unavailable, then an alternative, or approximate model is often used in its place.  In light of the well known adage of \cite{BD-1987}, ``All models are wrong, but some are useful", such models can nevertheless be useful for researchers to understand the fundamental relationships between variables. However, incorrect models may come with a non-negligible risk of leading to grossly misleading inference.\\

Specifying models through moment restrictions alone is appealing in many situations as a way to reduce the risk of misspecifying a likelihood function. The resulting class of models, commonly referred to as moment condition models, produces inference about parameters from the information supplied by the moment restrictions, thereby circumventing the need for potentially unsuitable distributional assumptions about the sampling density. Methods to estimate and conduct hypothesis tests based on moment condition models have been extensively developed in the Frequentist literature; for example the well-known generalized method of moments (GMM) (e.g. \citealp{Hansen-1982}, \citealp{NW-1987}, \citealp{Hall-2005} and many others). More recently, empirical likelihood (EL) based methods have become popular (e.g. \citealp{QL-1994}, \citealp{Imbens-1997}, \citealp{ISJ-1998}, \citealp{NS-2004} and \citealp{Schennach-2007}). Under a Bayesian framework, the estimation of moment condition models requires the formal construction of a proxy likelihood function using information given by the sample moments. See \cite{Yin-2009}, for example, who proposes the use of the asymptotic distribution of the sample moments as an approximation of the likelihood function -- an idea that is closely related to the use of a Laplace approximation as proposed by \cite{CH-2003}. In related work, and following on from earlier work by \cite{CI-2003}, Bayesian nonparametric methods are explored and further developed for moment condition models by \cite{BSS-2015}.\\

Bayesian EL-based methods build on theoretical work from the Frequentist perspective as initially developed by \cite{Owen-1988}. As argued by \cite{Lazar-2003}, the result is a valid substitute for the true likelihood function in a Bayesian analysis. In addition, \cite{Schennach-2005} shows that an exponentially tilted empirical likelihood (ETEL) has a well defined probabilistic interpretation arising from a Bayesian nonparametric procedure, and thus she proposes to obtain the Bayesian ETEL (BETEL) posterior density as being proportional to the product of the prior and the ETEL. Recently \cite{CSS-2017} propose a method under the BETEL framework to deal with the problem when some moment conditions are misspecified. They introduce additional nuisance parameters in a reformulation of the moment restrictions, resulting in new and valid so-called augmented moment conditions, enabling them to be used in a BETEL setting. Their approach then compares the models as defined by different moment conditions, and selects the model having the largest marginal likelihood, thus promoting inference about the parameters to be made from the moment restrictions consistent with having the most empirical support.\\ 

The focus of our work is to deal with the problem when outliers are present in the data while working within a moment condition model setting. Outlier-proned data has been a concern in econometrics and related disciplines for many years (see, for example the reviews given by \citealp{Stigler-1973}, \citealp{MMRK-1991} and \citealp{Berger-1994}). Outliers are described as observations that are not generated from the same DGP as the majority, and they have the potential to distort understanding of the fundamental relationship between most of the observations. Robust methods for moment condition models have been proposed under the Frequentist framework, for example \cite{RT-2001} and \cite{OT-2005robust} who consider robust estimation methods under the GMM setting. \cite{Schennach-2005} demonstrates in an example that moment conditions may be selected with robustness in mind, but does not elaborate further. Moment restrictions specified and valid for the majority of observations may in fact be invalid when outliers are present. \\ 

This paper contributes to the literature by proposing a novel robust Bayesian method based on the BETEL framework for moment condition models, intended for situations when outliers may be present in the data. Our new method, named the robust BETEL (RBETEL) method, is justified as the distribution that minimises the posterior expected loss function, along the lines of \cite{BHW-2016} and where the relevant loss function arises from an appropriately defined EL ratio (see, \citealp{Owen-1990}). Parameter values that result in a lower EL ratio correspond to stronger evidence for an hypothesis associated with the validity of given moment restrictions. The relevant moment conditions are specified in a manner similar to \cite{CSS-2017}, by introducing a nuisance indicator vector that is used to separate the complete set of observations into subsets of outlying and non-outlying data points. The new loss function for the RBETEL setting can be expressed as the EL ratio evaluated using only the non-outlying (or `active') data points given by the indicator vector. We then argue, using the framework of \cite{BHW-2016}, that the RBETEL joint posterior is a valid and coherent representation of subjective uncertainty about the minimizer of the expected loss. The robust posterior distribution is seen as the marginal RBETEL posterior for the model parameters, obtained using a Markov chain Monte Carlo (MCMC) approach, and marginalized over the uncertainty regarding the locations of active data points.\\ 

We demonstrate the estimation performance of the proposed RBETEL method under controlled simulation settings. The RBETEL method is shown to be robust to outliers in the sense that the posterior mean estimates of the model parameters are close to the (designed) parameter values used to simulate the `good' (i.e. non-outlying) data. In addition, we find that the percentage of the posterior densities whose mass covers cover the designed values is high. The RBETEL method is then used to analyze the relationship between the brain weight and body weight of various land animal species, using a dataset also used by \cite{RV-1990unmasking}, who apply a Frequentist approach to mitigate potential outliers and examine possible leverage points. A linear model is employed to estimate the relationship between the two variables following \cite{RV-1990unmasking}. Comparing the estimation results produced by both the BETEL and RBETEL methods, we find that the RBETEL method fits the majority of the observations well and seems to be robust to potential leverage effects. In addition, the posterior mean estimates of the parameters produced by the RBETEL method are similar to the parameter estimates produced by the robust M--estimator of \cite{Yohai-1987}.\\ 

The remainder of the paper is organized as follows. Section 2 provides background information, including a brief review of the original BETEL method of \cite{Schennach-2005} and the new framework that updates prior belief via a loss function proposed by \cite{BHW-2016}. Section 3 proposes the new RBETEL method, the loss function used for the new method is shown and the posterior distribution based on the loss function is derived. Details for the computation are also discussed. Simulation experiments are conducted in Section 4. Then in Section 5 the empirical study regarding the relationship between land animals' brain weight and body weight is provided. Section 6 concludes the paper and discusses the future work needed to improve the new method.

\section{Background} \label{s:Back}
This section provides essential background for the proposed robust Bayesian exponentially tilted empirical likelihood (RBETEL) method. We first review the standard Bayesian exponentially tilted empirical likelihood (BETEL) framework proposed by \cite{Schennach-2005}, which is designed to produce inference for a moment condition model. The original BETEL setting does not consider the situation when outliers are present in the data. We build upon the BETEL framework and develop the new RBETEL method in Section \ref{s:RBETEL}, which produces Bayesian inference that is robust to outliers that contaminate the observations. In addition, we outline the framework recently proposed by \cite{BHW-2016}. The posterior distribution is viewed as an update of prior belief via a loss function under this framework, and it provides an alternative justification for the BETEL posterior distribution. We identify a monotonic relationship between an empirical likelihood ratio and the loss function for the standard BETEL method, and thus a loss function for the new RBETEL method can be obtained in a similar way. The RBETEL posterior distribution can then be justified under the framework of \cite{BHW-2016}. 

\subsection{Bayesian exponentially tilted empirical likelihood}\label{s:BETEL}

A moment condition model is specified through a set of moment restrictions of the form
\begin{equation}\label{eq:BETEL_moment_condition}
	E^{F}\left[g(X;\theta)\right]=\mathbf{0},
\end{equation}
where $g(X;\theta)$ is a given $d_g$ dimensional function of a $d_X \times 1$ random vector $X$ and a $p \times 1$ parameter vector $\theta$. For identification purposes, $d_g \geq p$ must be satisfied. The expectation in (\ref{eq:BETEL_moment_condition}) is taken with respect to the unknown distribution of $X$, denoted by $F$, and $\mathbf{0}$ is a $d_g \times 1$ vector of zeros.\\ 

Suppose that a random sample $x_{1:n} = \left(x_1, \dots, x_n\right)$ is observed. Following \cite{Schennach-2005}, if the interior of the convex hull of $\bigcup_{i=1}^n\{g(x_i,\theta)\}$ contains the origin, then the BETEL posterior takes the form
\begin{equation}\label{eq:BETEL_tra_posterior}
	\pi_{\text{\tiny{BETEL}}}\left(\theta | x_{1:n}\right) \propto \pi(\theta) \widehat{p}(x_{1:n}|\theta),
\end{equation}
where $\pi(\theta)$ is the assumed prior probability density function (pdf) for $\theta$, and $\widehat{p}(x_{1:n}|\theta)$ is the proxy likelihood function, referred to as the ETEL. The ETEL is given by
\begin{equation}\label{eq:ETEL-tra}
	\widehat{p}(x_{1:n}|\theta) = \prod_{i=1}^{n}\widehat{w}_i(\theta),
\end{equation} 
where the $\widehat{w}_i(\theta)$, for $i=1,\dots,n$, are implied probabilities defined by the solution to the constrained minimization problem given by
\begin{equation}\label{eq:BETEL_tra_KLminimization}
	\left(\widehat{w}_1(\theta),\dots,\widehat{w}_n(\theta)\right) = \underset{\left(w_1,\dots,w_n \right)}{\arg\min} \sum_{i=1}^{n}  w_i\ln w_i,
\end{equation}
subject to
\begin{equation}\label{eq:BETEL_tra_constraints}
	\sum_{i=1}^{n}w_i = 1 \mbox{ and } \sum_{i=1}^{n}w_i g(x_i; \theta)= \mathbf{0}.
\end{equation}

As shown by \cite{Schennach-2005}, the implied probabilities in (\ref{eq:BETEL_tra_KLminimization}) correspond to those that minimise the Kullback-Leibler (KL) divergence from the approximating multinomial distribution associated with the probabilities $(w_1,\dots, w_n)$ to the empirical distribution having weights given by $(\frac{1}{n},\dots,\frac{1}{n})$. Further, these probabilities can be computed conveniently as
\begin{equation}\label{eq:BETEL_tra_weigths}
	\widehat{w}_i(\theta) = \frac{\exp\left(\widehat{\lambda}(\theta)' g(x_i;\theta)\right)}{\sum_{j=1}^{n}\exp\left(\widehat{\lambda}(\theta)' g(x_j;\theta)\right)}, 
	\mbox{ for } i=i,\dots, n, 
\end{equation}
where the so-called optimal `tilting' parameter, $\widehat{\lambda}(\theta)$, is given by
\begin{equation}\label{eq:BETEL_tra_lam}
	\widehat{\lambda}(\theta) = \underset{\lambda}{\arg\min} \sum_{j=1}^{n}\exp\left(\lambda' g(x_j;\theta)\right).
\end{equation}

\cite{Schennach-2005} also shows that the ETEL is equivalent to the limit of a non-parametric Bayesian procedure under a suitably defined non-informative prior. Therefore, as \cite{CSS-2017} point out, the BETEL method may be used to form the basis of a Bayesian semi-parametric analysis. In Section \ref{s:BHW} we justify the BETEL method from a different perspective.

\subsection{Updating prior belief via a loss function}\label{s:BHW}

In this section, we outline the recent work of \cite{BHW-2016} who propose a framework to update prior beliefs about parameters via a loss function. We then justify the BETEL method as such an updated distribution under this new framework.\\ 

Suppose we have a loss function $l(\theta,X)$ which is a function of a parameter $\theta$ and a random variable $X$ having an unknown distribution function $F.$ Let $\theta_0$ be defined as the minimizer of the expected loss, i.e.

\begin{equation}\label{eq:BHW_minimizer}
	\begin{split}
		\theta_0 &= \underset{\theta}{\arg \min } E^F\left[l(\theta;X)\right]\\ 
		&= \underset{\theta}{\arg \min }\int l(\theta;X)dF.
	\end{split}
\end{equation} 

The objective of \cite{BHW-2016} is to determine the form of a valid and coherent representation of subjective uncertainty in $\theta_0$, denoted by $\widehat{\pi}(\theta|x_{1:n})$, from a prior $\pi(\theta)$ and observations $x_{1:n}$.\\ 

\cite{BHW-2016} argue that, although $F$ is unknown, a coherent decision maker should prefer a probability measure $\pi_1(\theta|x_{1:n})$ over $\pi_2(\theta|x_{1:n})$ if the posterior expected loss under $\pi_1(\theta \mid x_{1:n})$ is strictly less than the posterior expected loss under $\pi_2(\theta \mid x_{1:n})$, i.e. if
\begin{equation}\label{eq:BHW_decision}
	\int\int l(\theta;x_{1:n})dF\ \pi_1(\theta|x_{1:n}) d\theta < \int\int l(\theta;x_{1:n})dF\ \pi_2(\theta|x_{1:n}) d\theta.
\end{equation}
They also argue that provided the sample is independent of the prior, then a valid and coherent representation of subjective uncertainty in $\theta_0$ arises from minimizing the so-called cumulative (or additive) loss, given by
\begin{equation}\label{eq:BHW_cumulative_loss}
	\begin{split}
		& L\left(\pi(\theta|x_{1:n}); \pi(\theta), x_{1:n}\right)\\ 
		&= \int l(\theta; x_{1:n}) \pi(\theta|x_{1:n}) d\theta + \int \pi(\theta|x_{1:n})\log\left(\frac{\pi(\theta|x_{1:n})}{\pi(\theta)}\right)d\theta\\
		&= \int \pi(\theta|x_{1:n})\log\left(\frac{\pi(\theta|x_{1:n})}{\exp\{ -l(\theta; x_{1:n})\}\pi(\theta)}\right)d\theta.
	\end{split}
\end{equation}
This cumulative loss represents the expected posterior loss associated with the data $x_{1:n}$ in addition to the expected loss due to the prior $\pi(\theta)$. \cite{BHW-2016} further show that $L\left(\pi(\theta|x_{1:n}); \pi(\theta), x_{1:n}\right)$ $ \rightarrow \int\int l(\theta;x_{1:n})dF \pi(\theta|x_{1:n}) d\theta$ as $n \rightarrow \infty$, providing an asymptotic justification for the finite sample minimizer in (\ref{eq:BHW_cumulative_loss}).\\ 

The cumulative loss function (\ref{eq:BHW_cumulative_loss}) has the form of the KL divergence between the posterior distribution $\pi(\theta|x_{1:n})$ and another distribution with probability density function (pdf) proportional to $\exp\{ -l(\theta, x_{1:n})\}\pi(\theta)$. Therefore, it is straightforward to see that the minimizer of (\ref{eq:BHW_cumulative_loss}) has the form
\begin{equation}\label{eq:BHW_posterior}
	\begin{split}
		\widehat{\pi}(\theta|x_{1:n}) 
		&= \underset{\pi(\theta|x_{1:n})}{\arg\min} L\left(\pi(\theta|x_{1:n}); \pi(\theta), x_{1:n}\right)\\ 
		&= \frac{\exp\{ -l(\theta; x_{1:n})\}\pi(\theta)}{\int \exp\{ -l(\theta; x_{1:n})\}\pi(\theta) d\theta}.
	\end{split}
\end{equation}

\subsubsection{The BETEL loss function}\label{s:BETEL_loss}
We now define the loss function implied by the BETEL method arising from the framework of \cite{BHW-2016}. Given the BETEL posterior in (\ref{eq:BETEL_tra_posterior}), with implied probabilities $\widehat{w}_i(\theta)$, for $i=1,2,\ldots, n$ given in (\ref{eq:BETEL_tra_weigths}), and optimal tilting parameter $\widehat{\lambda}(\theta)$ given in (\ref{eq:BETEL_tra_lam}), it follows directly that the corresponding loss function must be given by	
\begin{equation}\label{eq:BETEL_loss_function}
	l_{\text{\tiny{BETEL}}}(\theta,x_{1:n}) = -\sum_{i=1}^{n}\log\widehat{w}_i(\theta).
\end{equation}
The BETEL posterior (\ref{eq:BETEL_tra_posterior}) can then be interpreted as the representation of the subjective uncertainty in $\theta_0$.\\ 

Notice that the exponential of the negative loss function is proportional to the EL ratio, $\mathcal{R}(\theta)=\prod_{i=1}^{n} n\widehat{w}_i(\theta)$, since
\begin{equation}\label{eq:BETEL_loss_ELR_relation}
	\mathcal{R}(\theta) \propto \prod_{i=1}^{n} \widehat{w}_i(\theta) =\exp\left(-l_{\text{\tiny{BETEL}}}(\theta;x_{1:n})\right).
\end{equation} 
Under the Frequentist framework and given an estimate of $\theta$, say $\widehat{\theta}$, the EL ratio statistic, $\mathcal{R}(\widehat{\theta})$, is commonly used for testing the null hypothesis that the moment restrictions in (\ref{eq:BETEL_moment_condition}) are valid (see, for example \citealp{Owen-1988} and \citealp{Owen-1990}). The asymptotic properties of the EL ratio using implied probabilities defined by (\ref{eq:BETEL_tra_KLminimization}) are discussed by \cite{Schennach-2007}.\\ 

A higher EL ratio, or equivalently, a lower $l_{\text{\tiny{BETEL}}}(\theta;x_{1:n})$ in our case, suggests less evidence against the null hypothesis, suggesting that we will be interested in values of $\theta$ for which $\mathcal{R}(\theta)$ is high and $l_{\text{\tiny{BETEL}}}(\theta;x_{1:n})$ is low. This intuition provides a guideline for constructing the loss function in the RBETEL method proposed in Section {\ref{s:RBETEL}}.\\

\section{Robust Bayesian exponentially tilted empirical likelihood method} \label{s:RBETEL}
In this section, we develop the proposed RBETEL method. Our approach is to augment the required moment conditions with information about the possible presence of outliers in the dataset. After showing that the standard moment conditions which assume all the observations have the same distribution will be invalid when outliers are present in the data, we modify the moment conditions by introducing an indicator vector into the formulation. We then demonstrate the form of the RBETEL posterior based on these modified moment conditions. The RBETEL posterior distribution is justified under the framework of \cite{BHW-2016} with a loss function related to the empirical likelihood ratio evaluated using some subsets of the data. We detail the computation strategy to sample model parameters and the indicator vector from the RBETEL posterior. Lastly, we discuss the choice of the moment conditions which provide the essential information to separate the good data and outliers. 

\subsection{A contaminated sample}
Given that outliers are thought to be present in the sample, we think of the distribution $F$ from which the data are actually sampled as a mixture of two distinct distributions, denoted by $G$ and $B$, respectively. In particular, a subset of the complete dataset consisting of $K$ observations are generated from the unknown distribution $G$, where $\frac{n}{2} < K\leq n$. We refer to $G$ as the `good' distribution, as the values drawn from $G$ are expected to satisfy the stated moment conditions. The remaining $n-K$ observations are assumed to arise from another distribution denoted by $B$, whose values have the potential to be vastly different from those generated by $G$. In particular, the moment conditions specified for $G$ will not hold for $B$. We refer to the distribution $B$ in this context as the `bad' distribution. \\

The reason for the imposed condition $\frac{n}{2} < K \leq n$ will become clearer in Section \ref{s:RBETEL_post}. It relates to the fact that as the RBETEL procedure will use identified subsets of non-outlying datasets in order to construct a suitable likelihood function, it will be important to ensure that at least half of the data come from $G$.\\ 

When the outlying realisations from $B$ appear randomly throughout the sample, with each realization in the complete dataset having an equal probability $v$ of being drawn from $G$, then the distribution associated with the data, without conditioning on $K$, is given by the mixture distribution
\begin{equation}\label{eq:RBETEL_data_DGP}
	\begin{split}
		& X_i \mid v \stackrel{iid}{\sim}
		\begin{cases}
			& G \mbox{ with prob } v\\ 
			& B \mbox{ with prob } (1-v),
		\end{cases}
	\end{split}
\end{equation}
where $X_i$ is a $d_x\times 1$ random variable corresponding to the realized value $x_i$, for each $i=1,2,\ldots,n$, and $\textsl{iid}$ refers to independent and identically distributed. \\

The \textit{expected} proportion of outliers, given by $(1-v)$, may be treated as either fixed or known. By suitable incorporation of a prior distribution for $v$, uncertainty regarding the expected proportion of outliers may also be accounted for, as we show in Section \ref{s:RBETEL_computation}. Until then, however, we treat $v$ as fixed.\\

Our interest lies in understanding the fundamental relationship between the subset of random variables generated from $G$ alone, and \textit{not} from the contaminated distribution in (\ref{eq:RBETEL_data_DGP}). In particular we want to infer a distribution for the value of the parameter $\theta\in\Theta$ from the \textit{valid} moment conditions
\begin{equation}\label{eq:RBETEL_objective_moment}
	E^{G}\left[g(X,\theta)\right]=\mathbf{0},
\end{equation}
for some $g(X,\theta)$ satisfying the usual conditions, as detailed in Section \ref{s:BETEL}. Note that the expectation in (\ref{eq:RBETEL_objective_moment}) is taken with respect to $G$, i.e. the distribution that generates the `good' data. We assume that the moment conditions are valid under $G$, meaning that (\ref{eq:RBETEL_objective_moment}) holds, while at the same time
\begin{equation}\label{eq:RBETEL_objective_moment_BAD}	E^{B}\left[g(X,\theta)\right] \neq \mathbf{0},
\end{equation}
meaning that the same type of moment restrictions will be invalid under the distribution $B$ that generates the outliers. Therefore, since $F$ is comprised of both $G$ and $B$, the moment restrictions will also be invalid under the mixture distribution, $F=v*G + (1-v)*B$, as detailed in (\ref{eq:RBETEL_data_DGP}).

\subsection{The RBETEL modified moment conditions}\label{s:RBETEL_moments}

The mixture model for $X_i$ in (\ref{eq:RBETEL_data_DGP}) may be written as the marginal distribution that results from the joint specification of $X_i$ and an auxiliary random variable $s_i$ that indicates whether observation $i$ comes from the `good' or `bad' mixture component, i.e.
\begin{equation}\label{eq:RBETEL_model_DGPmix}
	X_i \mid s_i \sim
	\begin{cases}
		& G \mbox{ if } s_i=1\\ 
		& B \mbox{ if } s_i=0,
	\end{cases}
\end{equation}
where the indicators in the vector $s=(s_1,\dots,s_n)$ are, by virtue of the mixture model, $\textsl{iid}$ Bernoulli($v$) random variables, consistent with $Pr(s_i=1 \mid v)=v$.\\ 

Using the mixture indicator variable, $s_i$, the desired moment expression in (\ref{eq:RBETEL_objective_moment}) may be expressed as
\begin{equation}\label{eq:RBETEL_moment_condition_model}
	E^{F}\left[g(X_i,\theta)s_i\right]=\mathbf{0} \mbox{ for } i=1,\dots,n,
\end{equation}
where we note that the expectation is taken with respect to $F$ rather than $G$. Now, since 
\begin{equation} \label{eq:iterated_expectation}
	E^{F}\left[ g(X_i,\theta) s_i \right] = v E^{G}\left[ g(X_i,\theta) s_i \mid s_i=1 \right] + (1-v) E^{B} \left[ g(X_i,\theta)s_i \mid s_i=0 \right],
\end{equation}
then $E^{F}\left[g(X_i,\theta) s_i\right]=\mathbf{0}$ only when $E^{G}\left[g(X_i,\theta)\right]=\mathbf{0}$.
It is clear that if $X_i$ is an outlier, then the moment condition in (\ref{eq:RBETEL_moment_condition_model}) will only be valid if the indicator $s_i$ is zero.\\ 

\subsection{The RBETEL posterior} \label{s:RBETEL_post}

The RBETEL method produces a posterior distribution for the parameter $\theta$ \textit{jointly} with the vector of indicator variables $s=(s_1,s_2,\ldots,s_n)$. In order to ensure that the indicators reflect the non-outlying observations (i.e. observations from $G$ rather than from $B$) we impose the constraint that  $K=\sum_{i=1}^{n} s_i> \frac{n}{2}$ to ensure that the resulting posterior inference about $\theta$ is informed by the majority of the data. Thus, given a prior specification for $\theta$ (consistent with the distribution $G$), denoted by $\pi(\theta)$, the RBETEL joint posterior conditional on $v$ is given by 
\begin{equation}\label{eq:RBETEL_posterior_given_v}
	\pi_{\text{\tiny{RBETEL}}}(\theta,s \mid x_{1:n},v) \propto \pi(\theta) \pi(s\mid v) \widetilde{p}(x_{1:n}\mid \theta, s) I_{\left(\sum_{j=1}^{n}s_j> \frac{n}{2}\right)},
\end{equation}
where the ETEL in this context is denoted by $\widetilde{p}\left(x_{1:n}|\theta,s\right)$. The indicator function $I_{(\cdot)}$ takes value one if the constraint holds, otherwise it is equal to zero. Although not strictly required, we would anticipate that $\pi(\theta)$ is independent of both $s$ and $v$, as implied by (\ref{eq:RBETEL_posterior_given_v}).\\ 

For a given indicator vector $s$ we are able to identify the outlier observations, each corresponding to $s_i=0$, and hence we will want to exclude these observations from the calculation of the implied probability weights. Accordingly, we note that the indicator vector $s$ determines the vector of $n$ empirical probabilities, given by $(\frac{s_1}{\sum_{j=1}^{n}s_j},..., \frac{s_n}{\sum_{j=1}^{n}s_j}).$ The corresponding RBETEL implied probabilities, given by $\left(\widetilde{w}_1(\theta,s),\dots,\widetilde{w}_n(\theta,s)\right)$, are then obtained by minimizing the KL divergence from the multinomial probabilities $(w_1s_1,...,w_ns_n)$ to the empirical probabilities noted above. Hence
\begin{equation}\label{eq:RBETEL_KLminimization}
	\left(\widetilde{w}_1(\theta,s),\dots,\widetilde{w}_n(\theta,s)\right)=\underset{\left(w_1,\dots,w_n \right)}{\arg\min} \sum_{i=1}^{n} w_i s_i \ln \left(w_i\sum_{j=1}^{n}s_j\right),
\end{equation}
subject to
\begin{equation}\label{eq:RBETEL_Constraints}
	\sum_{i=1}^{n}w_i s_i = 1 \mbox{ and } \sum_{i=1}^{n} w_i s_i g\left(x_i, \theta\right) = 0.
\end{equation}
From (\ref{eq:RBETEL_Constraints}) it can be seen that the weights for the active observations must sum to one, and the associated weighted sample average must satisfy the required conditions, corresponding to the theoretical specification in (\ref{eq:RBETEL_moment_condition_model}).\\

It can be found from (\ref{eq:RBETEL_KLminimization}) that the values of the implied probabilities, $\widetilde{w}_i(\theta,s)$, associated with $s_i=0$, are not uniquely defined. This does not cause any issue in the RBETEL setting, because these implied probabilities corresponding to outliers do not contribute to the value of the ETEL, given by 
\begin{equation}\label{eq:RBETEL_ETEL}
	\widetilde{p}\left(x_{1:n} \mid \theta,s\right) = \prod_{i=1}^{n} \left(\widetilde{w}\left(\theta,s\right) \sum_{j=1}^{n}s_j\right)^{s_i}.
\end{equation} 
It is straightforward to show that the RBETEL implied probabilities corresponding to the unit indicators, i.e. $s_i=1$, can be represented by \\
\begin{equation}\label{eq:RBETEL_weigths}
	\widetilde{w}_i(\theta,s)=\frac{\exp(\widetilde{\lambda}(\theta, s)'g(x_i, \theta))}{\sum_{j=1}^{n}s_j\exp(\widetilde{\lambda}(\theta, s)'g(x_j, \theta))},
\end{equation}
where the optimal tilting parameter is given by
\begin{equation}\label{eq:RBETEL_lam}
	\widetilde{\lambda}(\theta, s) = \underset{\lambda}{\arg\min} \sum_{j=1}^{n}s_j\exp(\lambda'g(x_j, \theta)).
\end{equation}

Therefore, although there are actually infinitely many solutions for the RBETEL weights in (\ref{eq:RBETEL_weigths}) given $s$ and $\theta$, the collection of the $K$ weights $\{\widetilde{w}_i(\theta,s)$ for all $i=1,2,...,n$,  where $s_i=1 \} $ is unique if the usual requirements on the functions comprising $g(X,\theta)$ and the parameter space $\Theta$ are satisfied.\\ 

Now, it is clear that if the indicator vector, $s$, is known, the RBETEL posterior is the same as a standard BETEL posterior conditional on a subset of the complete dataset. Hence for any $s$ and $\theta$ (given $v$) the joint RBETEL posterior pdf can be computed up to a normalizing constant, as per (\ref{eq:RBETEL_posterior_given_v}). However, to extract inference about $\theta$ \textit{marginal} of $s$, we simply marginalise the joint RBETEL posterior over $s$ to obtain
\begin{equation}\label{eq:RBETEL_posterior_theta}
	\pi_{\text{\tiny{RBETEL}}}(\theta \mid x_{1:n},v) \propto \int \pi_{\text{\tiny{RBETEL}}}(\theta,s \mid x_{1:n},v) ds.
\end{equation}  
As we show in Section \ref{s:RBETEL_computation}, computation of the posterior is undertaken using an MCMC approach, and hence this marginalisation will be easy to implement once a sample of $(\theta,s)$ draws from the joint posterior is available.   

\subsection{Posterior computation with MCMC}\label{s:RBETEL_computation}

In this section, we detail an MCMC method for sampling the unknowns $\theta$ and $s$ from the RBETEL joint posterior distribution, corresponding to (\ref{eq:RBETEL_posterior_given_v}). In addition, as uncertainty in the proportion of outliers, $v$ can be accommodated by hierarchically adding a prior distribution for $v$, in which case the \textit{joint} RBETEL posterior becomes
\begin{equation}\label{eq:RBETEL_posterior}
	\pi_{\text{\tiny{RBETEL}}}(\theta,s,v \mid x_{1:n}) \propto \pi(\theta) \pi(v) \pi(s \mid v) I_{\left(\sum_{j=1}^{n}s_j> \frac{n}{2}\right)}\prod_{i=1}^{n} \left(\widetilde{w}\left(\theta,s\right) \sum_{i=1}^{n}s_i\right)^{s_i},
\end{equation}
with, as before, the marginalisation to $\pi_{\text{\tiny{RBETEL}}}(\theta \mid x_{1:n})$ being managed via MCMC.\\

The approach we use here samples values of $\theta$, $s$ and $v$, iteratively, each from the corresponding full conditional RBETEL posterior. The sampling methods for $\theta$ and $v$, respectively, are reasonably straightforward. Both are undertaken conditionally on $s$ and hence effectively operate as BETEL conditional distributions on an active sub-sample. We detail these two cases first, followed by a description of our approach to sample $s$. 

\subsubsection{Sampling $\theta$ and $v$}

Only the currently active data subset is needed for sampling $\theta$, as $\pi_{\text{\tiny{RBETEL}}}(\theta \mid x_{1:n},v,s)=\pi_{\text{\tiny{RBETEL}}}(\theta \mid x_{1:n},s)$. This full conditional distribution is given by
\begin{equation}\label{eq:Sample_theta}
	\pi_{\text{\tiny{RBETEL}}}(\theta \mid x_{1:n},s) \propto 
	\pi(\theta) \prod_{i=1}^{n}\left(\widetilde{w}_i(\theta,s)\sum_{i=1}^{n}s_i\right)^{s_i}.
\end{equation}
As the $\widetilde{w}_i(\theta,s)$ must be calculated numerically, we suggest sampling from (\ref{eq:Sample_theta}) using a random walk Metropolis-Hastings (MH) algorithm (as described, for example, in \citealp{CG-1995MH}). \\

Ideally, good subjective prior information will be available for the proportion of outliers, $v$. Noting that the Beta distribution is conjugate for the Binomial distribution, and hence for the independent Bernoulli latent variables contained in $s$, we suggest the use of a suitably truncated Beta prior for $v$. In our examples we take the truncation to be $v \in (0.5,1]$, and use a truncated Beta prior for the parameter $v$, having hyperparameters $\alpha_0$ and $\beta_0$ and with pdf given by
\begin{equation}\label{eq:Prior_v}
	\pi(v) = \frac{v^{\alpha_0-1}(1-v)^{\beta_0-1}}{\int_{0.5}^{1}v^{\alpha_0-1}(1-v)^{\beta_0-1}dv}I\left(v\in(0.5,1]\right).
\end{equation}
We of course acknowledge that other priors may be used, if desired. However, under the truncated Beta prior specified in (\ref{eq:Prior_v}), the full conditional distribution for $v$ has a pdf given by
\begin{equation}\label{eq:Sample_v}
	\pi(v \mid x_{1:n},\theta,s)
	\propto v^{\sum_{i=1}^{n}s_i+\alpha_0-1}(1-v)^{n-\sum_{i=1}^{n}s_i+\beta_0-1} I(v\in (0.5,1 ]).
\end{equation}
Note that this full conditional does not depend on either $\theta$ or $x_{1:n}$, but corresponds to a kernel of a truncated Beta distribution with parameters given by $\sum_{i=1}^{n}s_i+\alpha_0$ and $n-\sum_{i=1}^{n}s_i+\beta_0$, with the value of $v$ restricted to $\left(0.5,1 \right]$.

\subsubsection{Sampling $s$}
Next, note that the indicators $s_i$, for $i=1,\dots, n$, are assumed \textit{a priori} to be \textit{iid} Bernoulli random variables with the probability of success given by $v$.  This implies that the prior distribution for $K=\sum_{j=1}^{n}s_j$ is $Bernoulli(v)$. However, once the data are observed, the conditional joint posterior distribution for $K$ will no longer be $Bernoulli(v)$ and the individual $s_j$ components will no longer be \textit{iid}. In addition, recall that we have imposed the identifying constraint that $K>\frac{n}{2}$ so that we will always include the majority of the observations in the `good', or `active' data group. To overcome these complications, we use an MH approach to sample $s$. To devise an efficient candidate proposal, however, we do so by sampling a joint draw of the pair $(s, K)$, accepting with the relevant probability, and then discarding the drawn value of $K$.\\ 

Noting the form of the joint full conditional distribution of $s$ and $K$ is given by
\begin{equation}\label{eq:fullcond_s_K}
	\pi_{\text{\tiny{RBETEL}}}(s,K \mid x_{1:n},\theta, v) \propto
	v^{K}(1-v)^{n-K}I_{(K>\frac{n}{2})}I_{(K=\sum_{j=1}^ns_j)} \prod_{i=1}^{n}\left(\widetilde{w}_i(\theta,s)K\right)^{s_i}.
\end{equation}
We denote the joint proposal of $(s,K)$ at the $m^{th}$ iteration of the MCMC procedure by $(s^{\dagger},K^{\dagger}).$ The approach we take is to first generate $K^{\dagger}$, conditionally upon $s^{(m-1)}$, and then generate $s^{\dagger}$ given both $K^{\dagger}$ and $s^{(m-1)}$. In particular, only $K^{(m-1)}=\sum_{j=1}^{n}s_j^{(m-1)}$ is used when generating $K^{\dagger}$. Our proposal distribution is therefore given by
\begin{equation}\label{eq:Proposed_s_two_step}
	q_{(s,K)}(s^{\dagger},K^{\dagger}\mid s^{(m-1)})
	= q_2\left(s^{\dagger}\mid K^{\dagger}, s^{(m-1)}, \tau\right) 
	q_1\left(K^{\dagger}\mid K^{(m-1)}\right),
\end{equation}
where $\tau$ is a tuning parameter and is defined below. The proposal distribution (\ref{eq:Proposed_s_two_step}) is comprised of two parts. The first part, denoted by $q_1(\cdot)$, proposes a new value for the sum of the indicators, $K$, conditional on its value in the previous MCMC iteration, $K^{(m-1)}$. We use a truncated $Binomial\left(n,\frac{c}{n}K^{(m-1)}\right)$ here, with $c=0.99$ used in this paper, so that $K^{\dagger}$ will be similar to $K^{(m-1)}$ while ensuring that there is always a positive probability of moving away from $K^{(m-1)}=n$ if that ever occurs. The proposed value $K^\dagger$ must be integers in the interval $(\frac{n}{2}, n]$.\\

The second part of the proposal involves drawing the candidate indicator vector of $s^{\dagger}$, conditionally given the drawn $K^{\dagger}$ and the last MCMC draw $s^{(m-1)}$. For this second proposal we select the vector $s^{\dagger}$ such that exactly $K^{\dagger}$ of its components have $s_i^{\dagger}=1$, indicating the set of active observations from the complete dataset in the next iteration of the MCMC. The probability of an observation to stay in the same state, i.e. active with $s^{(m-1)}=1$ or inactive with $s^{(m-1)}=0$, as proportional to some fixed probability, say $\tau$, while the probability of $s_i$ changing to the alternative state given by $1-\tau$. All this while conditioning on the total number of proposed active observation being fixed at $K^{\dagger}$. Accordingly, the pdf corresponding to the second proposal component is
\begin{equation}
	\begin{split}
		q_2\left(s^{\dagger}\mid K^{\dagger}, s^{(m-1)}, \tau\right) \propto  &\prod_{\{i:s_i^{(m-1)}=1\}} \left[ \tau^{s_i^{\dagger}}(1-\tau)^{(1-s_i^{\dagger})} \right]\\
		&\prod_{\{i:s_i^{(m-1)}=0\}} \left[(1 - \tau)^{s_i^{\dagger}}\tau^{(1-s_i^{\dagger})} \right]I_{(\sum s_i=K^\dagger)}.
	\end{split}
\end{equation}

At iteration $m$, the generated indicator vector proposal, $s^{\dagger}$, is then accepted, activating a new subset of data, with probability given by 
\begin{equation}\label{eq:MH_ratio_indicator}
	R_{s,K} = \min\left(1, \frac{\pi_{\text{\tiny{RBETEL}}}(s^{\dagger},K^{\dagger} \mid x_{1:n},\theta, v)}{\pi_{\text{\tiny{RBETEL}}}(s^{(m-1)},K^{(m-1)} \mid x_{1:n},\theta, v)}
	\frac{q_{s,K}(s^{(m-1)},K^{(m-1)})\mid s^{\dagger})} {q_{s,K}(s^{\dagger}, K^{\dagger}\mid s^{(m-1)})}\right),
\end{equation}
otherwise, set $(s^{(m)},K^{(m)})=(s^{(m-1)},K^{(m-1)}).$ Note that as $K^{(m)}$ may be determined completely from $s^{(m)}$, only $s^{(m)}$ need actually be retained.\\



\subsection{The RBETEL loss function}\label{s:RBETEL_loss}

We now provide a justification for the RBETEL posterior distribution using the framework of \cite{BHW-2016} that we explained in Section \ref{s:BHW}. After identifying a loss function for the RBETEL method, the joint posterior distribution for $(\theta, s)$ taking the form (\ref{eq:RBETEL_posterior_given_v}) is shown to be a representation of subjective uncertainty in the values of $\theta$ and $s$ which minimize the expected RBETEL loss. We note that this derivation is undertaken conditionally on $v$.\\ 

Consistent with the algebraic derivation and intuition provided in Section {\ref{s:BETEL_loss}}, the RBETEL loss function is related to an EL ratio defined by 
\begin{equation}\label{eq:RBETEL_ELR}
	\begin{split}
		\widetilde{\mathcal{R}}(\theta,s) &= \frac{\prod_{i=1}^{n}\left(s_i\widetilde{w}_i(\theta,s)\right)^{s_i}}{\prod_{i=1}^{n}\left(\frac{s_i}{\sum_{j=1}^{n}s_j}\right)^{s_i}}\\ 
		&= \prod_{i=1}^{n}\left(\widetilde{w}_i(\theta,s)\sum_{j=1}^{n}s_j\right)^{s_i}.
	\end{split}
\end{equation}

The EL ratio function (\ref{eq:RBETEL_ELR}) can be written equivalently as 
\begin{equation}\label{eq:RBETEL_ELR_equivalent}
	\widetilde{\mathcal{R}}(\theta,s)=\prod_{i=1}^n \frac{\widetilde{w}_i(\theta,s)}{1/\sum_{j=1}^{n}s_j} I(s_i=1),
\end{equation}
which is the EL ratio evaluated using only the active observations. This EL ratio is higher when evaluated on values of $\theta$ and $s$ that are consistent with the desired theoretical moment condition (\ref{eq:RBETEL_moment_condition_model}). Information provided by the moment conditions regarding the distinguishing characteristics of the good data and outliers enters the RBETEL posterior through this EL ratio.\\ 

Now, the RBETEL loss function, $\widetilde{l}(\theta,s;x_{1:n})$, is defined following the monotonic relationship (\ref{eq:BETEL_loss_ELR_relation}) found in Section \ref{s:BETEL_loss}, i.e.
\begin{equation}\label{eq:RBETEL_loss_function}
	\begin{split}
		\widetilde{l}\left(\theta,s;x_{1:n}\right) &= -\ln \widetilde{\mathcal{R}}(\theta,s)\\ 
		&= -\sum_{i=1}^{n}s_i\ln \left(\widetilde{w}_i(\theta,s) \sum_{j=1}^{n}s_j\right).
	\end{split}
\end{equation}

Following \cite{BHW-2016}, we derive the RBETEL joint posterior distribution, $\pi_{\text{\tiny{RBETEL}}}(\theta,s \mid x_{1:n},v)$, with loss function (\ref{eq:RBETEL_loss_function}) in the appendix. The RBETEL joint posterior distribution is obtained from minimization of a cumulative loss function which is a finite sample version of the expect loss, $E^F\left[\widetilde{l}(\theta,s; x_{1:n})\right]$. 

\subsection{Choosing the moment conditions}\label{s:Choice_Moment}

It is essential to carefully choose the moment restrictions which provide information about the relationship between variables and the distribution of the data. In the literature, where the presence of outliers in the dataset has not been considered, moment conditions are chosen from available model assumptions, for example that the conditional expectation of an error term should equal zero, or that certain products of random variables should, on average, be zero. These are commonly referred to as orthogonality assumptions (see e.g., \citealp{Hansen-1982}, \citealp{wooldridge-2001-applyGMM} and \citealp{SWY-2002}). Such choices are at the discretion of the analyst, and in general will be problem specific.\\ 


Now that outliers are to be accommodated, we assume that the analyst has already in mind some moment conditions, such as those arising from orthogonality conditions for the `good' data. These conditions are now stated as $E^G[g(\theta,x)]=\mathbf{0}$. However, the RBETEL method also requires that the selected moment condition vector $E^F[g(\theta,x)] \neq \mathbf{0}$ when outliers are present, i.e. when $E^B[g(\theta,x)] \neq \mathbf{0}$. Therefore, it is important to include some additional moment conditions in the specification to ensure this additional need is met. In fact, as discussed in Section \ref{s:BHW}, the key information for identifying outliers comes from the moment conditions that lead to different values of the EL ratio when evaluated using different subsets of observations. Under the RBETEL framework, at least $p$ moment conditions are needed to ensure that the parameter vector $\theta$ is identified. Among the restrictions used, at least one moment condition will need to be chosen to fit the characteristics of the good data while simultaneously being misspecified for the outliers. For convenience, we refer to such a moment condition as a `\textsl{key condition}'. More than one key condition may be included, and in fact more than one will help to improve the identification of the outliers, as these key conditions ensure that the specified set of RBETEL moment conditions are valid only under $G$, and not under $B$.\\

However, since the true DGP is unknown, the key conditions will need to be built from the analyst's understanding of the characteristics of the good data, as well as how these characteristics might be destroyed by outliers. Again these considerations will be problem specific, however we offer a few suggestions for location and regression problems below.\\ 

\subsubsection{Suggested key conditions for a location problem}
Suppose one is interested in inferring the location, $\mu$, of a population associated with the `good' part of an observed dataset, $(x_1,\dots,x_n)$, which may be contaminated by outliers. A commonly used moment condition for inferring the location of $G$ is $E^G\left[x - \mu \right]=0$. However, in the RBETEL context, the moment condition given by $E^F\left[(x - \mu)s \right]=0$ alone would not be sufficient because it does not provide information to distinguish observations from $G$ and $B$.\\  

One approach would be to augment the simple moment condition $E^F\left[(x - \mu)s \right]=0$ above with an additional `key condition' arising from a certain assumption about $G$. For example, it might be expected that
\begin{equation*}
	E^G[(x_i-\mu)^3]=0,
\end{equation*}
which would lead to the RBETEL additional specification that
\begin{equation*}
	E^F[(x_i-\mu)^3s_i]=0.
\end{equation*}
This key condition suggests that the good data distribution should be symmetric around $\mu$, so that observations that cause substantial skewness in the complete dataset may be considered as outliers.\\ 

A second possible approach to the specification of key conditions in the location problem is to consider moment conditions derived from a robust Huber-type location estimator, given by $g(x,\theta)= H(x-\mu)$ where for some fixed $\epsilon_0>0,$ the $H(\cdot)$ function is given by
\begin{equation}\label{eq:Huber_H}
	H(\epsilon)= 
	\begin{cases}
		& 1 \mbox{ if } \epsilon \geq \epsilon_0,\\ 
		& \epsilon/\epsilon_0 \mbox{ if } \left|\epsilon \right| < \epsilon_0,\\
		& -1 \mbox{ if } \epsilon \leq -\epsilon_0.
	\end{cases}
\end{equation}
A key condition in this case could be formed according to
\begin{equation*}
	E^{F}[\left((x_i-\mu) - H(x_i-\mu)\right)s_i]=0.
\end{equation*}

A third alternative would be to incorporate selected statistics known for their (Frequentist) robust properties in to the moment conditions. An example we consider is the key condition given by
\begin{equation*}
	E^{F}\left[(x_i-\mu)^2s_i \right] - MAD(x_{1:n})^2 = 0,
\end{equation*}
where $MAD(x_{1:n})$ denotes the median absolute deviation of the observations. This key condition incorporates a robust scale estimate of the data to provide information regarding the expected (squared) distance between good data points and the conditional mean value $\mu$.\\ 

\section{Simulation experiments} \label{s:Sim}
In this section, we demonstrate the performance of the RBETEL method under controlled simulation settings. Section 4.1 examines the RBETEL method to estimate location using the moment conditions, introduced in Section \ref{s:Choice_Moment}, with a simulated dataset. Then we generate replicated datasets from the location estimation setting with a fixed proportion of outliers in Section 4.2. We examine the average performance of the RBETEL posterior inference when outliers in the data have different sizes. The standard BETEL method is also applied to estimate the location of the data and the inferential results produced by both methods are compared. In Section 4.3, replicated datasets are considered in a linear regression setting when the data may be contaminated by outliers. Some of these outliers are leverage points which are known to have strong influence on inference in the linear regression context. Under this setting, we consider different numbers of outliers in the generated datasets.

\subsection{Location estimation using different key conditions}
Here we consider a location estimation example and compare the estimated RBETEL posterior distributions when different key conditions suggested in Section \ref{s:Choice_Moment} are included into the set of moment conditions.\\ 

We simulate a dataset $y_{1:n}=(y_1,\dots,y_n)$ from the following data generating process (DGP):
\begin{equation}\label{eq:Sim1_DGP}
	y_i =
	\begin{cases}
		1 + e_i, \mbox{ with probability } 0.95,\\ 
		6 + e_i, \mbox{ with probability } 0.05,
	\end{cases}
\end{equation}
where the errors, $e_i$, are independently generated from a Normal distribution with zero mean and variance equal to one, i.e. $e_i\overset{iid}{\sim}N(0,1)$, for $i=1,\dots,n$. One hundred observations are generated (i.e. $n=100$), with $95\%$ of these having a mean of $1$, and these observations are referred to as the good data points. The remaining $5\%$ of the observations are considered to be outliers, and they have a mean of $6$. Only a single dataset is used here to present a clear picture of how the entire posterior is impacted by the RBETEL conditions. Simulation experiments that employ replicated datasets are considered in Sections \ref{s:Sim_Location} and \ref{s:Sim_Linear}.\\ 

\begin{figure}[h]
	\centering
	\includegraphics[width=0.6\textwidth]{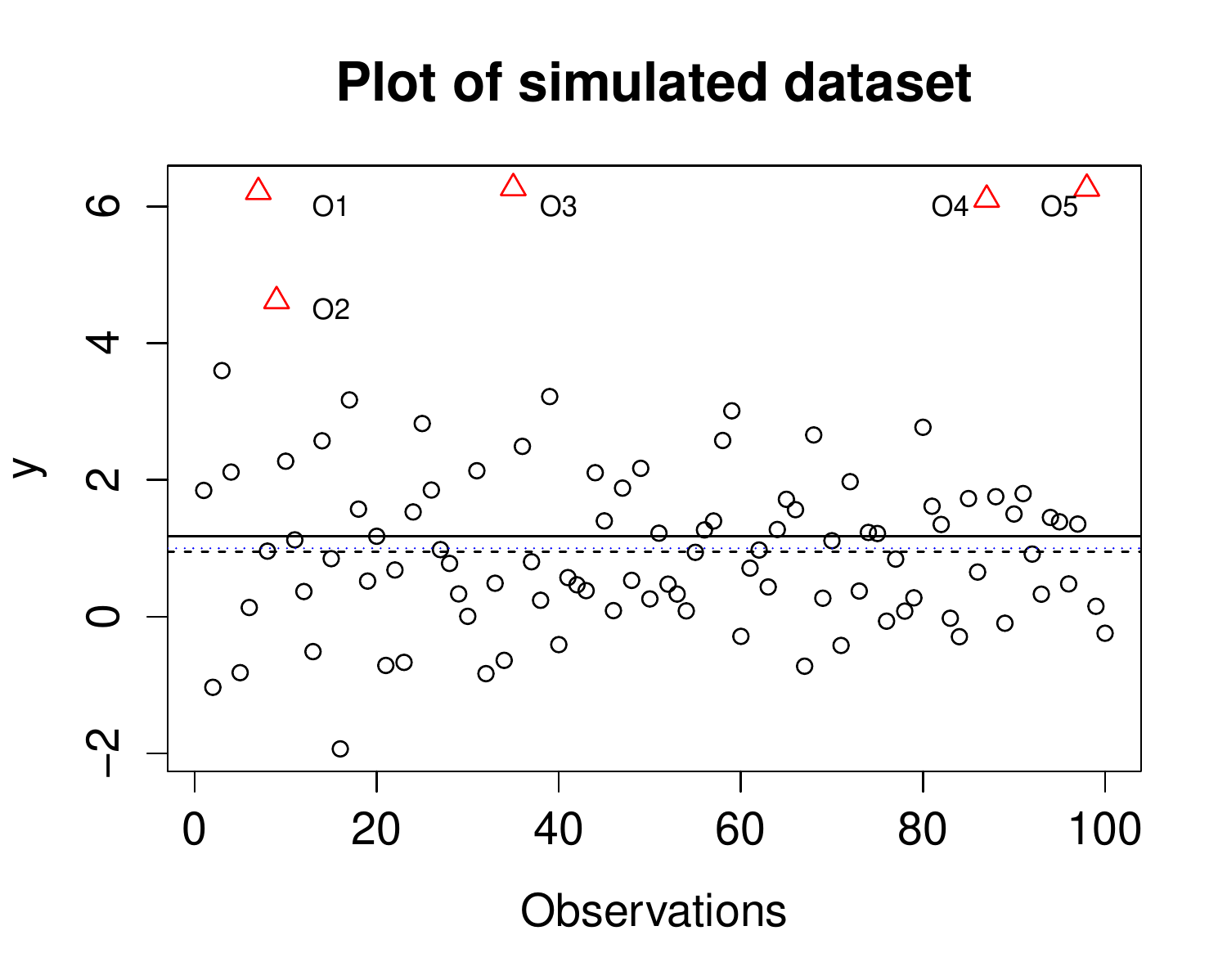}
	\caption{Simulated dataset with $5\%$ of the observations being outliers. The outliers are highlighted with red triangles. The solid black line indicates the sample mean of the data ($1.17$). The dashed line is the median of the data ($0.95$). The dotted blue line indicates the value $1$ which is the designed location for the good data.}
	\label{fig:SimOneData}
\end{figure}

The simulated dataset is plotted in Figure \ref{fig:SimOneData} and five outliers ($O_1=6.21$, $O_2=4.62$, $O_3=6.27$, $O_4=6.10$, $O_5=6.26$) in the simulated data are highlighted. We can see that the sample mean, which is about $1.17$, deviates from the designed location for the good data at $1$. On the other hand, the median of the data, which is about $0.95$, is very close to the desired value of $1$. In addition, the designed DGP produces outliers with expected location much larger than the good data. As a result, the complete dataset seems to have an asymmetric distribution and this differs from the distribution of the good data which is symmetric.\\ 

We consider producing posterior distributions for the location of the good data, denoted by $\mu$, using the RBETEL method based on different set of moment conditions. The moment condition $E\left[(y_i - \mu)s_i\right]$ is always used, together with some combinations of the following key conditions:  
\begin{equation*}
	\mbox{C1: } E\left[(y_i - \mu)^3s_i\right],
\end{equation*} 
\begin{equation*}
	\mbox{C2: } E[\left((y_i-\mu) - H(y_i-\mu)\right)s_i]=0,
\end{equation*}
\begin{equation*}
	\mbox{C3: } E\left[(y_i-\mu)^2s_i \right] - MAD(y_{1:n})^2 = 0.
\end{equation*}
The function $H(\cdot)$ in $C2$ is given by (\ref{eq:Huber_H}) and the trimming parameter, $\epsilon_0$, is set to be $1.5$.
$MAD(y_{1:n})$ in $C3$ denotes the median absolute deviation of the observations.\\ 

The BETEL method is also applied to this simulated dataset and inference is based on a single moment condition $E\left[y-\mu\right]=0.$\\ 

We produce $30,000$ MCMC draws, discarding the first $10,000$ draws and keeping the subsequent $20,000$ draws for posterior inference. We assign a flat Normal prior with mean equal to zero and variance equal to $100$ to the location parameter $\mu$, i.e. $\mu \sim N(0,100)$. The priors for the indicator vector, $s$, and the probability parameter, $v$, are the same as those discussed in Section 3.4.1. The hyperparameters for $\pi(v)$, given in (\ref{eq:Prior_v}), are set to $\alpha_0=50$ and $\beta_0=5$.\\

The inferential results produced by the standard BETEL and RBETEL methods are summarized in Table \ref{tab:BETEL_OneData_Sim} and Table \ref{tab:RBETEL_OneData_Sim}, respectively. We report posterior means (Post.Mean), posterior standard deviations (Post.SD) and time-series standard errors (TS.SE) for the location parameter, $\mu$, produced by both methods. For the RBETEL method, we also report the expected proportion of good data, $v$. In addition, we report the probabilities of each outlier being used in the MCMC iterations; each of these probabilities can be interpreted as the probability of the corresponding observation being a good data point.\\

The inferential results show that the RBETEL method produces robust inference for the location parameter $\mu$ based on moment conditions that include any combination of key conditions $C1$, $C2$ and $C3$. The posterior means are all close to $1$ which is the designed location of the good dataset. On the other hand, the posterior mean estimate for parameter $\mu$ given by the standard BETEL method is $1.2148$, which is near the sample mean of the complete dataset. The posterior standard deviations and the time-series standard errors produced by the RBETEL method with different moment conditions are also similar.\\

\begin{figure}[h]
	\centering
	\includegraphics[width=0.6\textwidth]{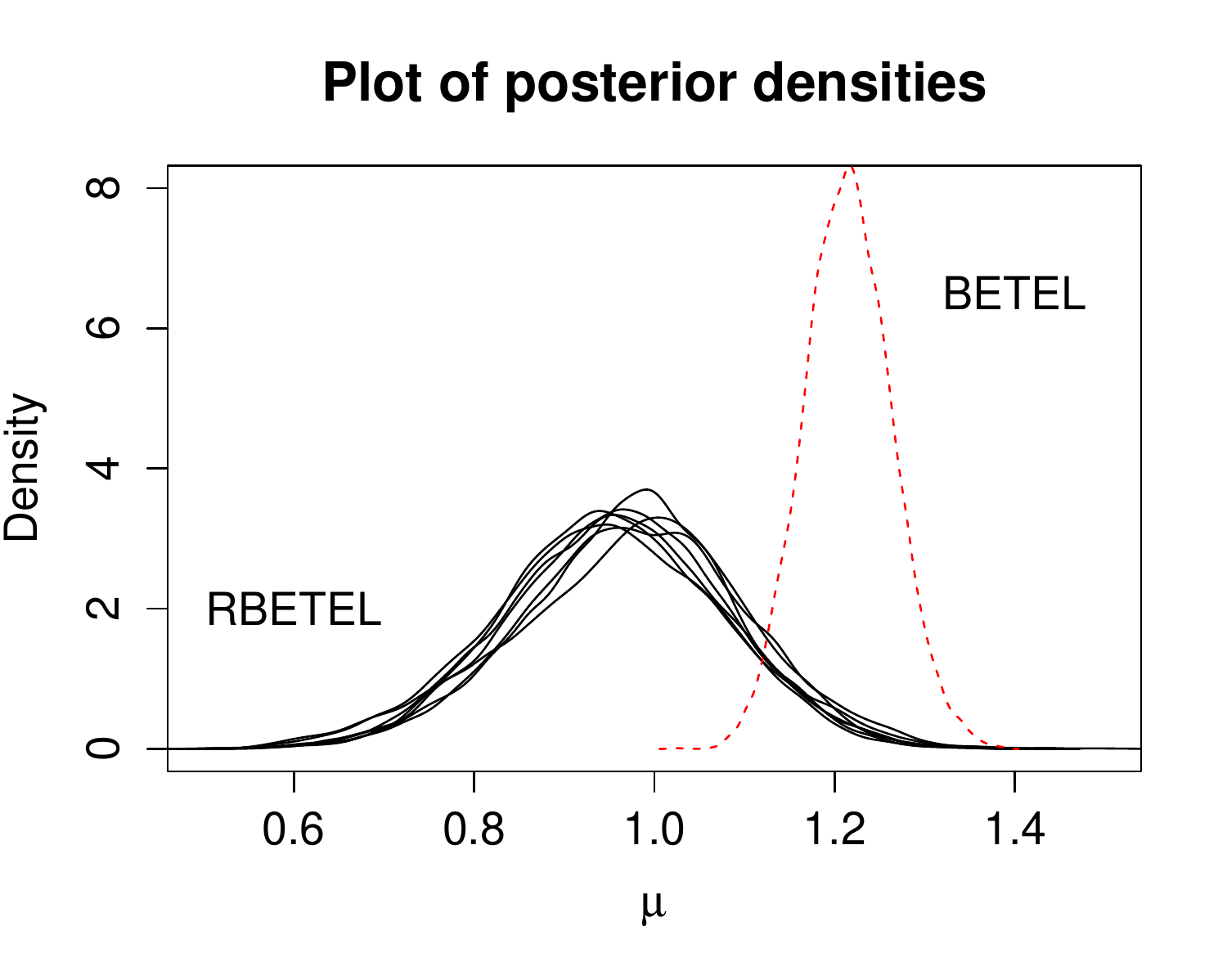}
	\caption{Posterior densities of $\mu$ produced by the standard BETEL method (red dashed curve), and the RBETEL methods (black solid curve) based on different moment conditions.}
	\label{fig:OneData_PostDensity_Sim}
\end{figure}

Figure \ref{fig:OneData_PostDensity_Sim} shows the kernel estimates of the posterior densities for $\mu$ produced by standard BETEL method, and the RBETEL methods based on different moment conditions. The posterior densities produced by the RBETEL method all center near the designed value $1$, while the BETEL posterior does not cover the designed value.\\ 

The estimated probabilities of the outliers, $O_1$, $O_2$, $O_3$, $O_4$ and $O_5$, being treated as good data (shown in Table \ref{tab:RBETEL_OneData_Sim}) are small in all the cases, while the corresponding probabilities for good data points are all larger than $90\%$. We find that these probabilities are smaller when condition $C_1$, which suggests that the good data have a symmetric distribution, is included into the set of moment conditions. This suggests that a key condition that provides correct information about the distribution of the good data seems to improve the performance of the RBETEL method.\\

\begin{table}[H]
	\centering
	\begin{tabular}{lrrr}
		\cline{2-4}    \multicolumn{1}{r}{} & \multicolumn{1}{l}{Post.Mean} & \multicolumn{1}{l}{Post.SD} & \multicolumn{1}{l}{TS.SE} \\
		\hline
		$\mu$ & 1.2148 & 0.0485 & 0.0025 \\
		\hline
	\end{tabular}%
	\caption{BETEL posterior summaries for the location parameter given the simulated dataset.}
	\label{tab:BETEL_OneData_Sim}%
\end{table}%

\begin{table}[H]
	\begin{adjustbox}{max width=\textwidth}
		\fontsize{11}{11}\selectfont
		\centering
		\begin{tabular}{ll|rrr|rrrrr}
			\cline{3-10}          & \multicolumn{1}{r}{} & \multicolumn{1}{l}{Estimates} &       &       & \multicolumn{2}{l}{$Pr(S_i=1)$} &    &       &  \\
			\cline{3-10}          & \multicolumn{1}{r}{} & \multicolumn{1}{l}{Post.Mean} & \multicolumn{1}{l}{Post.SD} & \multicolumn{1}{l|}{TS.SE} & \multicolumn{1}{l}{$O_1$} & \multicolumn{1}{l}{$O_2$} & \multicolumn{1}{l}{$O_3$} & \multicolumn{1}{l}{$O_4$} & \multicolumn{1}{l}{$O_5$} \\
			\hline
			$C1$  & $\mu$ & 0.95 & 0.12 & 0.0047 & 0.01 & 0.10 & 0.01 & 0.02 & 0.02 \\
			& $v$   & 0.90 & 0.03 & 0.0004 &       &       &       &       &  \\
			\hline
			$C2$  & $\mu$ & 0.98 & 0.12 & 0.0051 & 0.09 & 0.14 & 0.09 & 0.09 & 0.08 \\
			& $v$   & 0.90 & 0.03 & 0.0003 &       &       &       &       &  \\
			\hline
			$C3$  & $\mu$ & 0.96 & 0.12 & 0.0045 & 0.19 & 0.23 & 0.17 & 0.18 & 0.17 \\
			& $v$   & 0.91 & 0.03 & 0.0003 &       &       &       &       &  \\
			\hline
			$C1 \&C2$ & $\mu$ & 0.97 & 0.12 & 0.0049 & 0.05 & 0.16 & 0.04 & 0.05 & 0.05 \\
			& $v$   & 0.90 & 0.03 & 0.0003 &       &       &       &       &  \\
			\hline
			$C1 \& C3$ & $\mu$ & 0.94 & 0.13 & 0.0051 & 0.03 & 0.11 & 0.03 & 0.03 & 0.03 \\
			& $v$   & 0.90 & 0.03 & 0.0004 &       &       &       &       &  \\
			\hline
			$C2 \& C3$ & $\mu$ & 0.96 & 0.12 & 0.0047 & 0.10 & 0.14 & 0.10 & 0.09 & 0.09 \\
			& $v$   & 0.90 & 0.03 & 0.0003 &       &       &       &       &  \\
			\hline
			$C1 \& C2 $ & $\mu$ & 0.97 & 0.13 & 0.0057 & 0.07 & 0.19 & 0.06 & 0.07 & 0.06 \\
			$\& C3$ & $v$   & 0.90 & 0.03 & 0.0003 &       &       &       &       &  \\
			\hline
		\end{tabular}%
	\end{adjustbox}
	\caption{RBETEL posterior summary results based on different moment conditions. The left panel reports posterior mean, posterior standard deviation and time-series standard error for parameters $\mu$ and $v$. The right panel shows the probability of each outlier being included in the sub-samples for analysis.}
	\label{tab:RBETEL_OneData_Sim}%
\end{table}%

\subsection{Inference regarding location under different outlier sizes} \label{s:Sim_Location} 
In this section, we examine the performance of the RBETEL inference when outliers in the dataset have different sizes. The RBETEL inferential results are also compared with the estimates produced by the standard BETEL method, using replicated datasets.\\ 

The DGP we consider is given by
\begin{equation*}
	y_i =
	\begin{cases}
		\mu_0 + e_i, \mbox{ with probability } 0.95,\\ 
		\xi_0 + e_i, \mbox{ with probability } 0.05,
	\end{cases}
\end{equation*}
where $\mu_0$ is the mean for $95\%$ of the observations which are good data, and the other $5\%$ of observations are treated as outliers and they have a mean equal to $\xi_0$. The errors are independently generated from a Normal distribution with zero mean and variance equal to one, i.e. $e_i\overset{iid}{\sim}N(0,1)$, for $i=1,\dots,n$.\\ 

We simulate $100$ datasets and each consists of $n=1000$ observations. The mean for the good data is set to be $\mu_0=1$ for all the simulated datasets. We consider cases when the mean of the outliers, $\xi_0$, is equal to $2$, $4$ and $6$, which correspond to small, medium and large outlier sizes, respectively.\\ 

We use all three key conditions examined in the previous section as moment conditions for RBETEL approach and the moment conditions are given by
\begin{equation}\label{eq:Sim1_moment_condition_RBETEL}
	\begin{split}
		& E\left[(y_i-\mu)s_i\right] = 0\\ 
		& E\left[(y_i - \mu)^3s_i\right] = 0 \mbox{ \ for } i=1, \dots, n,\\
		& E[\left((y_i-\mu) - H(y_i-\mu)\right)s_i]=0,\\ 
		& E\left[(y_i - \mu)^2\ s_i\right] - MAD(y_{1:n})^2 = 0. \\ 
	\end{split}
\end{equation}

It is clear that the data characteristics in (\ref{eq:Sim1_moment_condition_RBETEL}) are not suitable for the complete dataset when outliers are present. This implies that when applying the standard BETEL method, $E^F\left[y-\mu \right]=0$ is the only moment condition we can use to estimate $\mu$.\\ 

We produce $30,000$ MCMC draws throughout the simulation experiments, discarding the first $10,000$ draws and keeping the subsequent $20,000$ draws for posterior inference. We assign a flat Normal prior with a mean equal to zero and a variance equal to $100$ to the location parameter $\mu$, i.e. $\mu \sim N(0,100)$. The priors for the indicator vector, $s$, and the probability parameter, $v$, are the same as those discussed in Section 3.4.1. The hyperparameters for $\pi(v)$, given by (\ref{eq:Prior_v}), are set to $\alpha_0=500$ and $\beta_0=50$.\\

The inferential results are summarized in Table \ref{tab:Simulation1}. We report the average posterior means (Av.Post.Mean), average posterior standard deviations (Av.Post.SD) and average time series standard errors (Av.TS.SE). In addition, we look at the proportion of coverage (P.O.C) which is the proportion of the $95\%$ posterior credible intervals (C.I.), out of $100$ simulated datasets, that cover the designed location of the good data, i.e. $\mu_0=1$.\\

We can see that the results produced by the standard BETEL method are influenced by outliers in the data. The BETEL posterior means for datasets with small, medium and large outliers are given by $1.0531$, $1.1545$ and $1.2505$, respectively. These posterior mean estimates are all near the sample means associated with the corresponding complete datasets. When the size of outliers is small, $92\%$ of the standard BETEL posteriors still cover the designed location of the good data. The influence of outliers is not so obvious, although on average the posterior means for $\mu$ indeed deviate from the designed value of $\mu_0=1$. However, it becomes more apparent that the BETEL method is non-robust with respect to outliers when the size of the outliers gets large. The BETEL posterior mean for the location $\mu$ deviates further from the designed value when the average size of the outliers is increased. When medium and large sized outliers are present in the data, none of the $95\%$ C.I.s cover the designed mean.\\

The newly proposed RBETEL method performs well regardless of the size of the outliers, in the sense that the average posterior means are near the designed location for the good data $\mu_0=1$. All of the $95\%$ credible intervals cover the desired value in all the cases. The average RBETEL posterior means for datasets with small, medium and large sized outliers are given by $1.0086$, $1.0095$ and $1.0071$, respectively. In fact, we have not found any evidence suggesting that the performance of RBETEL posterior mean estimator is affected the by average size of the outliers.\\ 

\begin{figure}[H]
	\centering
	\includegraphics[width=0.6\textwidth]{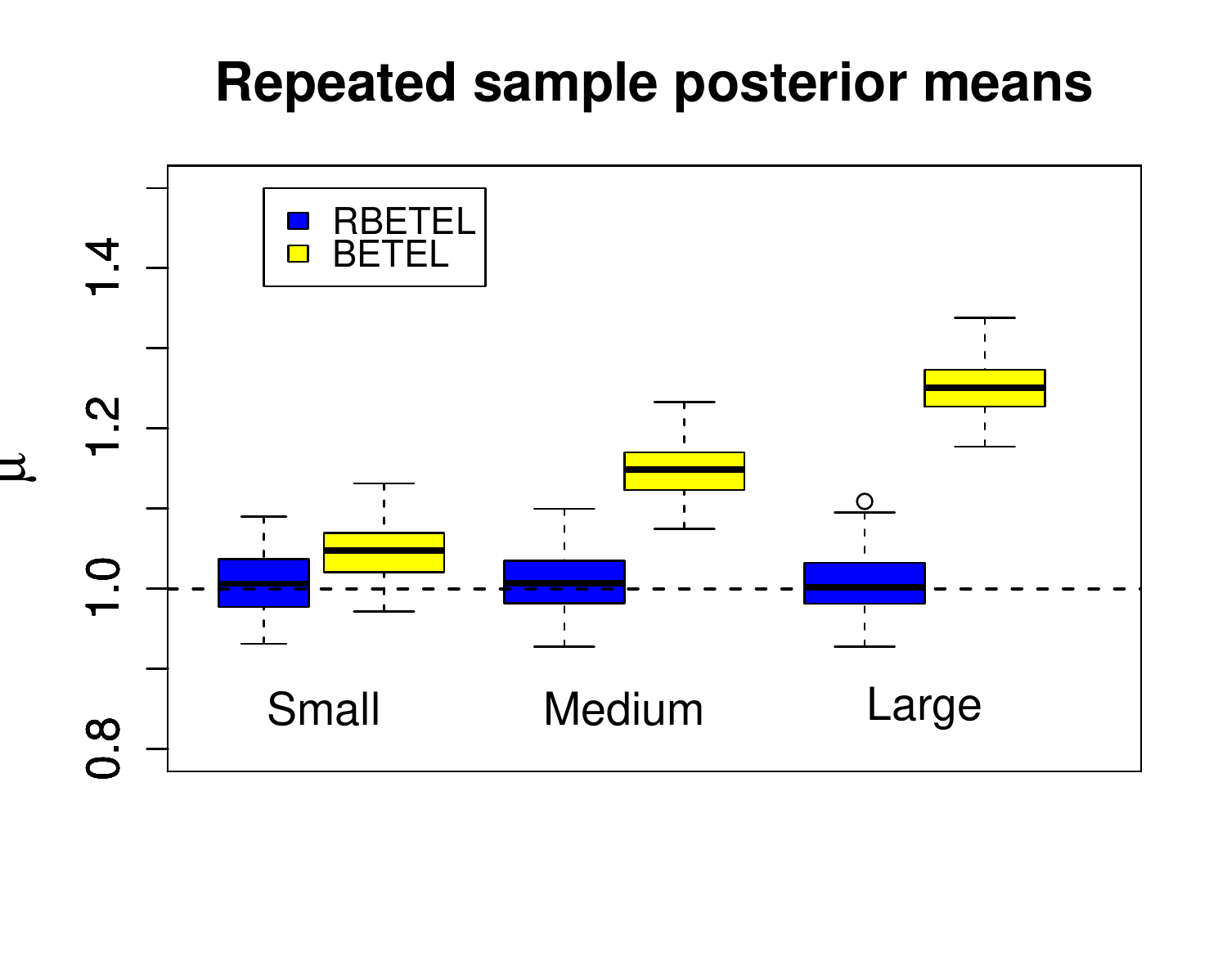}
	\caption{Boxplots of the posterior means produced by the RBETEL and standard BETEL methods. The boxplot on the left (blue)/right(yellow) of each pair corresponds to the posterior means given by the RBETEL/BETEL method. The horizontal line indicates the designed location for the good data, $\mu_0=1$.}
	\label{fig:Sim_ReapDataBox}
\end{figure}

We show boxplots of the posterior means produced by both methods for simulated data with different sized outliers in Figure \ref{fig:Sim_ReapDataBox}. We can see that the posterior means produced by the RBETEL method are concentrated around the designed location of good data for all cases. On the other hand, the posterior means given by standard BETEL method deviate further from the mean of the good data as the average size of outliers increases.\\ 

We also find that the posterior standard deviations produced by the RBETEL method are larger than the BETEL posterior standard deviations, which suggests that the RBETEL posteriors are on average more defused than the posteriors produced by the BETEL method under this simulation setting. This is mainly because the RBETEL posterior for $\mu$ is obtained by marginalizing over the posterior distributions conditional on subsets of the observations, with the proportion of outliers being unknown. It is also possible that this uncertinty may be offset by sharper posterior inference provided by the uncontaminated `good' data. However, this possibility does not appear to have occurred here. We will return to this point again in Section \ref{s:Sim_Linear}.\\  

\begin{table}[H]
	\centering
	\begin{tabular}{r|rrrr}
		\cline{2-5}    \multicolumn{1}{r}{} &       & \multicolumn{1}{l}{BETEL} &       &  \\
		\hline
		\multicolumn{1}{l|}{Size} & \multicolumn{1}{l}{Av.Post.Mean} & \multicolumn{1}{l}{Av.Post.SD} & \multicolumn{1}{l}{Av.TS.SE} & \multicolumn{1}{l}{P.O.C} \\
		\hline
		Small     & 1.0531 & 0.0653 & 0.0016 & 0.92 \\
		Median  & 1.1545 & 0.0663 & 0.0015 & 0 \\
		Large  & 1.2505 & 0.0671 & 0.0016 & 0 \\
		\hline
		
		\multicolumn{1}{r}{} &       &       &       &  \\
		\cline{2-5}    \multicolumn{1}{r}{} &       & \multicolumn{1}{l}{RBETEL} &       &  \\
		\hline
		\multicolumn{1}{l|}{Size} & \multicolumn{1}{l}{Av.Post.Mean} & \multicolumn{1}{l}{Av.Post.SD} & \multicolumn{1}{l}{Av.TS.SE} & \multicolumn{1}{l}{P.O.C} \\
		\hline
		Small     & 1.0086 & 0.1352 & 0.0058 & 1 \\
		Median  & 1.0095 & 0.1373 & 0.0051 & 1 \\
		Large  & 1.0071 & 0.1377 & 0.0055 & 1 \\
		\hline
	\end{tabular}%
	\caption{Estimation results for $\mu$ using BETEL and RBETEL methods. The reported values are average posterior mean, average posterior standard deviation, average time series standard error and the proportion of $95\%$ C.I.s that cover the designed value $\mu_0=1$.}
	\label{tab:Simulation1}
\end{table}%

\subsection{Linear regression with different proportions of outliers}\label{s:Sim_Linear}

In this section we detail the results of a separate simulation experiment where we consider inferring a linear relationship between variables when outliers may be present in the data. We examine the situations when different proportions of observations are treated as outliers. A few outliers in the simulated data are designed to be leverage points and these outliers are designed so that they have strong influence on non-robust inference. This way the difference between RBETEL inference and that of the standard BETEL method can be seen more easily.\\ 

We generate the independent variable, $x_i$, from a Normal distribution with a mean equal to zero and variance equal to $5$, i.e. 
\begin{equation*}
	x_i \overset{iid}{\sim} N(0, 5) \mbox{ for } i=1,\dots,n.
\end{equation*}
After sorting the independent variables in an ascending order, the dependent variables, $y_{1:n}=(y_1,\dots,y_n)$, are simulated by firstly generating
\begin{equation}\label{eq:Sim2_DGP1}
	y^\triangle_i =
	\begin{cases}
		\delta^*_0 + \delta^*_1x_i + e_i \mbox{ with probability } v^*,\\ 
		\delta^*_0 + \delta^*_1x_i + 3e_i, \mbox{ with probability } 1-v^*,
	\end{cases}
\end{equation}
and then setting
\begin{equation}\label{eq:Sim2_DGP2}
	y_i =
	\begin{cases}
		y^\triangle_i \mbox{ for } i=1,\dots,n-3,\\ 
		y^\triangle_i - 10, \mbox{ for } i=n-2, n-1, n.
	\end{cases}
\end{equation}
The errors $e_i$ are generated independently from a Normal distribution with mean equal to zero and standard deviation equal to one, i.e. $e_i\overset{iid}{\sim} N(0, 1)$, for $i=1,\dots,n$. The expected proportion of observations that are good data is given by $v^*$. The other observations are generated with error standard deviation equal to $3$, and they are considered to be outliers. The parameters in the DGP (\ref{eq:Sim2_DGP1}) with superscript `star' denote the designed values, and we set $\delta^*_0=2$ and $\delta^*_1=1$. We generate $n=1000$ observations for each simulated dataset, and reduce the last three observations by $10$ units so that they are leverage points.\\

\begin{figure}[H]
	\centering
	\includegraphics[width=0.6\textwidth]{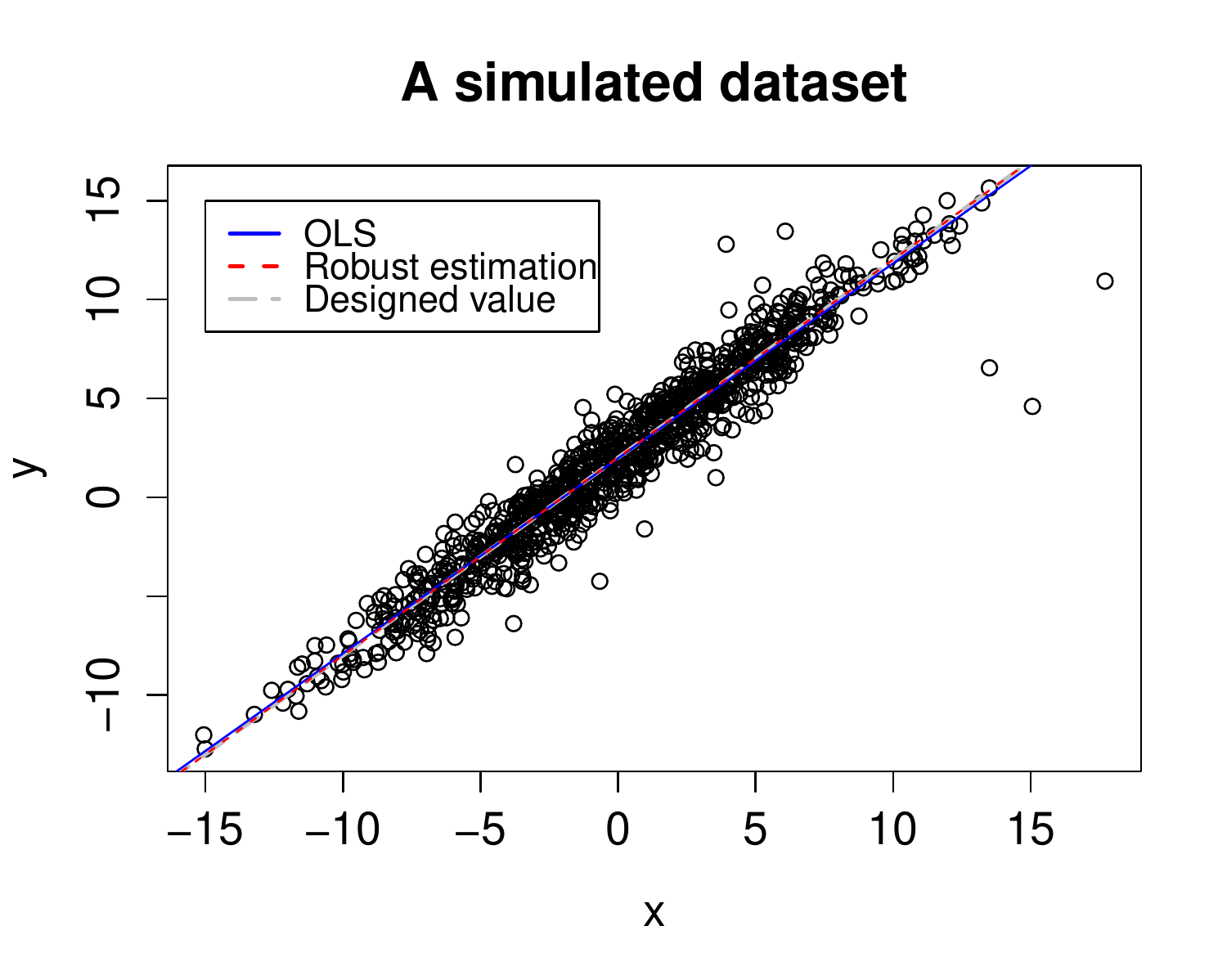}
	\caption{A simulated dataset with $5\%$ outliers and some leverage points. The solid blue line is estimated by OLS. The dashed red line is obtained from robust regression. The dotted black line is the designed regression line.}
	\label{fig:SimLinearOut}
\end{figure}

We plot a simulated dataset with $v^*=0.95$ in Figure \ref{fig:SimLinearOut}. We can see that that there are a few outliers in this data, and the leverage points on the right of the plot cause the ordinary least square (OLS) estimation of the regression line to have a smaller slope comparing to the designed value. A Frequentist robust regression inferential method proposed by \cite{Yohai-1987} is also employed to estimate the regression line, and it can recover the designed regression line very well in this case. The regression line estimated by the robust method basically overlaps with the designed regression line in the plot.\\ 

Commonly used moment conditions for estimating the linear regression model, when outliers are not considered, are given by
\begin{equation}\label{eq:Sim2_BETEL_moments}
	\begin{split}
		E\left[y_i-\delta_0-\delta_1x_i\right] = 0\\ 
		E\left[(y_i-\delta_0-\delta_1x_i)x_i\right] = 0.
	\end{split}
\end{equation}
We apply the standard BETEL method to estimate the linear regression relationship of the simulated data based on the moment conditions (\ref{eq:Sim2_BETEL_moments}).\\ 

We also include the basic moment conditions (\ref{eq:Sim2_BETEL_moments}) for the RBETEL inferential method, and these moment conditions are reformulated as
\begin{equation}\label{eq:Sim2_RBETEL_moment1}
	\begin{split}
		& E\left[(y_i-\delta_0-\delta_1x_i)s_i\right] = 0,\\ 
		& E\left[(y_i-\delta_0-\delta_1x_i)x_is_i\right] = 0.
	\end{split}
\end{equation}
In addition, we add some key conditions constructed following the suggestions in Section \ref{s:Choice_Moment}.\\ 

The first key condition is based on the symmetric distribution assumption for the error term, and it is given by
\begin{equation}\label{eq:Sim2_key1}
	E\left[(y_i-\delta_0-\delta_1x_i)^3s_i\right] = 0.
\end{equation}
This key condition may be violated when a subset of data contain outliers, in particular, when the outliers are asymmetrically distributed around the designed linear regression line.\\ 

Another set of key conditions is built incorporating a Huber-type location estimator for the error term. These key conditions are given by
\begin{equation}
	\begin{split}
		& E\left[\left((y_i-\delta_0-\delta_1x_i)-H(y_i-\delta_0-\delta_1x_i) \right)s_i\right]=0,\\ 
		& E\left[\left((y_i-\delta_0-\delta_1x_i)x_i-H(y_i-\delta_0-\delta_1x_i)x_i \right)s_i\right]=0,
	\end{split}
\end{equation}
where Huber's function $H(\cdot)$ is given in (\ref{eq:Huber_H}) and we set the trimming parameter $\epsilon_0$ to be $1.5$.\\ 

We construct the last RBETEL key condition for linear regression by incorporating a robust scale for the error term. We notice that the distinguishing characteristics of the outliers and the good data in the linear regression case is that an outlier is further away from the robust regression line than a good data point. A robust scale estimator for the errors can be seen as a summary statistic of the data that is related to the squared vertical distance of the good data to a regression line. We employ the robust regression methodology proposed by \cite{Yohai-1987} and denote the robust scale estimate for the error term by $T(x_{1:n}, y_{1:n})$, and then the key condition is given by
\begin{equation}\label{eq:RBETEL_linear_key}
	E\left[\left((y_i-\delta_0-\delta_1x_i)^2 - T(x_{1:n}, y_{1:n})\right)s_i\right]=0.
\end{equation}

We examine the cases when $v^*$ equals $1$, $0.98$, $0.95$ and $0.92$ in our experiments. $100$ replications of datasets are simulated from the designed DGP for each $v^*$. For both methods considered here, we produce and discard the first $20,000$ MCMC draws and keep the subsequent $30,000$ draws for inference. A flat Normal prior with a mean equal to zero and a variance equal to $100$ is assigned to the parameters $\delta_0$ and $\delta_1$. When applying the RBETEL method, the priors for $s$ and $v$ are specified according to Section \ref{s:RBETEL_computation}. The hyperparameters for $\pi(v)$ are set to $\alpha_0=500$ and $\beta_0=50$.\\ 

The Bayesian inferential results for parameters $\delta_0$ and $\delta_1$ are summarized in Table \ref{tab:Simulation2}. Both BETEL and RBETEL methods estimate the parameters accurately when the datasets do not contain outliers. The average posterior means are close to the desired values and all the posterior $95\%$ C.I.s cover the designed values for the parameters. When outliers are present in the data, we can see that the posterior means produced by the standard BETEL method deviate from the designed values and the proportions of $95\%$ C.I.s that cover the designed parameter values drop. In particular, the percentage of coverage drops to around $35\%$ for the slope parameter, $\delta_1$, because of the effect of leverage points. On the other hand, the RBETEL method shows robustness with respect to outliers. The RBETEL average posterior means for the parameters are close to the designed values that generate the good data and the percentages of posterior $95\%$ C.I.s covering the desired values are all higher than $97\%$ regardless of the proportion of outliers in the data.\\ 

\begin{figure}[H]
	\centering
	\begin{subfigure}[b]{0.45\textwidth}
		\includegraphics[width=\textwidth]{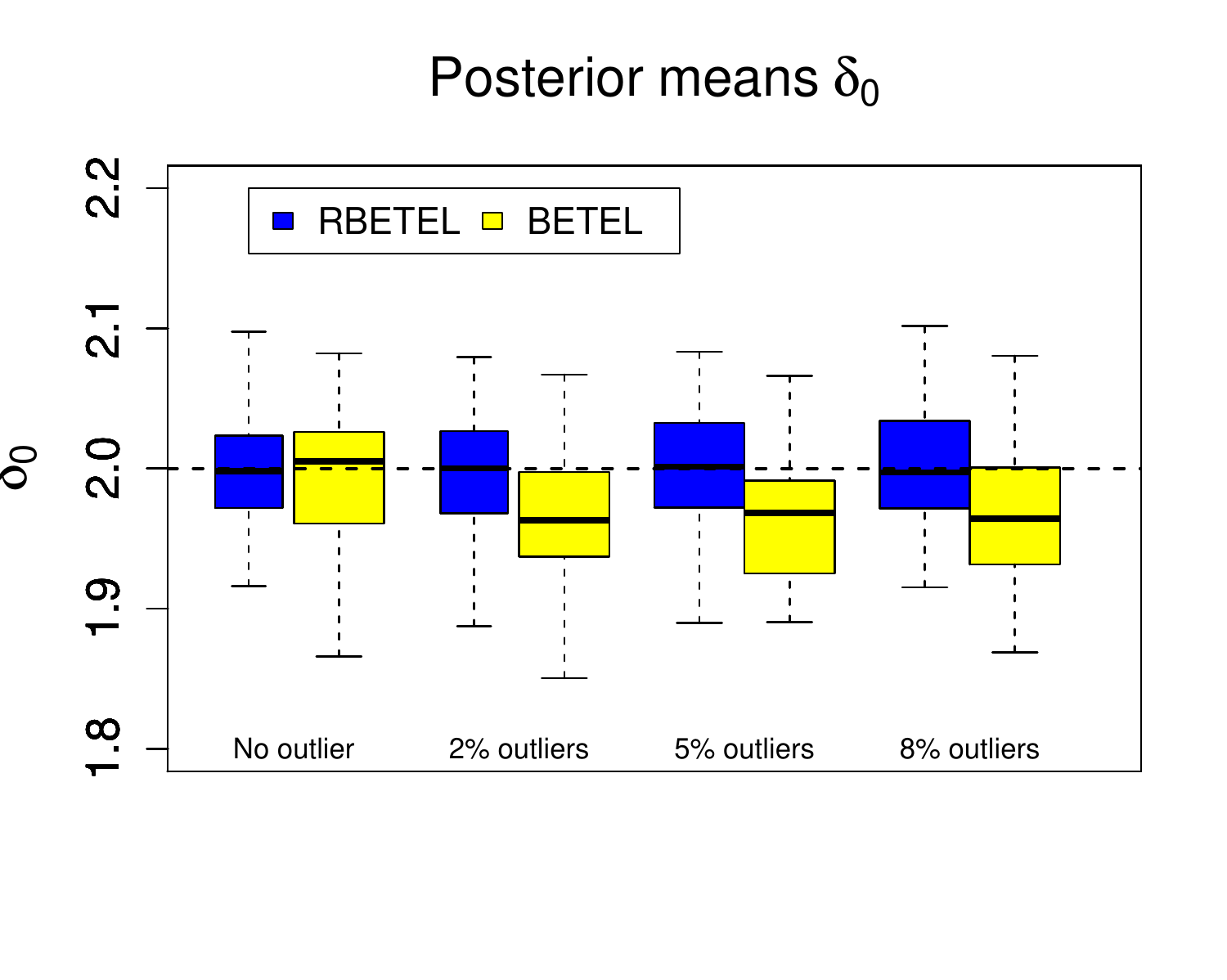}
		\caption{ }
		\label{fig:d0}
	\end{subfigure}
	\begin{subfigure}[b]{0.45\textwidth}
		\includegraphics[width=\textwidth]{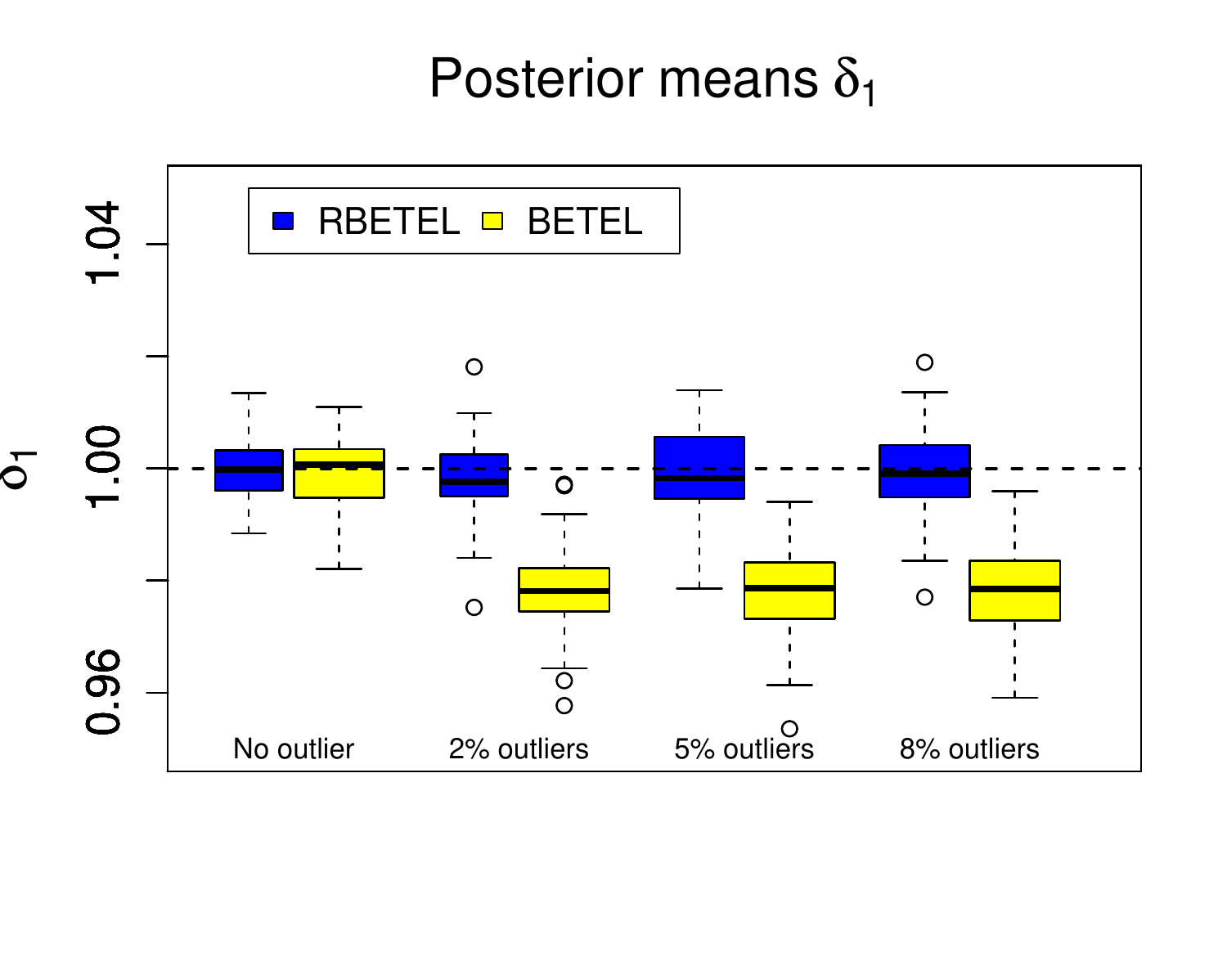}
		\caption{ }
		\label{fig:d1}
	\end{subfigure}
	\caption{Boxplots of the posterior means for the parameters $\delta_0$ and $\delta1$. The boxplot on the left (blue)/right(yellow) of each pair corresponds to the posterior means given by the RBETEL/BETEL method. The horizontal line indicates the designed parameter values for the good data.}
	\label{fig:Sim_Linear_Box}
\end{figure}

Figure \ref{fig:Sim_Linear_Box} shows the boxplots of the posterior means produced by the BETEL and RBETEL methods for the linear regression coefficients $\delta_0$ and $\delta_1$. In all cases, the posterior means produced by the RBETEL method for both parameters are concentrated around the designed parameter values. However, the posterior means produced by the standard BETEL method are not centered around the designed values when outliers are present. In particular, the BETEL posterior means for the slope coefficient $\delta_1$ are always below the designed value due to the effect of leverage points. In this setting, we also note that the average posterior standard deviation under the RBETEL method is actually slightly lower when compared against that from the BETEL method, highlighting the fact that trimming out contaminants from the dataset may actually sharpen the resulting posterior.\\ 

\begin{sidewaystable}
	\centering
	\begin{tabular}{r|rrrr|rrrr}
		\cline{2-9}    \multicolumn{1}{r}{} &       &       &       & \multicolumn{1}{l}{BETEL} &  &       &       &  \\
		\cline{2-9}    \multicolumn{1}{r}{} &       & \multicolumn{1}{l}{$\delta_0$} &       &       &       & \multicolumn{1}{l}{$\delta_1$} &       &  \\
		\hline
		\multicolumn{1}{l|}{$v^*$} & \multicolumn{1}{l}{\small Av.Post.Mean} & \multicolumn{1}{l}{\small Av.Post.SD} & \multicolumn{1}{l}{\small Av.TS.SE} & \multicolumn{1}{l|}{\small P.O.C} & \multicolumn{1}{l}{\small Av.Post.Mean} & \multicolumn{1}{l}{\small Av.Post.SD} & \multicolumn{1}{l}{\small Av.TS.SE} & \multicolumn{1}{l}{\small P.O.C} \\
		\hline
		1     & 2.00245 & 0.03185 & 0.00195 & 1 & 0.99975 & 0.00633 & 0.00014 & 1 \\
		0.98  & 1.96481 & 0.03827 & 0.00289 & 0.82 & 0.97791 & 0.01265 & 0.00049 & 0.34 \\
		0.95  & 1.96470 & 0.04100 & 0.00324 & 0.82 & 0.97712 & 0.01281 & 0.00048 & 0.35 \\
		0.92  & 1.96703 & 0.04429 & 0.00375 & 0.78 & 0.97742 & 0.01298 & 0.00049 & 0.34 \\
		\hline
		\multicolumn{1}{r}{} &       &       &       & \multicolumn{1}{r}{} &       &       &       &  \\
		\cline{2-9}    \multicolumn{1}{r}{} &       &       &       & \multicolumn{1}{l}{RBETEL} &       &       &       &  \\
		\cline{2-9}    \multicolumn{1}{r}{} &       & \multicolumn{1}{l}{$\delta_0$} &       &       &       & \multicolumn{1}{l}{$\delta_1$} &       &  \\
		\hline
		\multicolumn{1}{l|}{$v^*$} & \multicolumn{1}{l}{\small Av.Post.Mean} & \multicolumn{1}{l}{\small Av.Post.SD} & \multicolumn{1}{l}{\small Av.TS.SE} & \multicolumn{1}{l|}{\small P.O.C} & \multicolumn{1}{l}{\small Av.Post.Mean} & \multicolumn{1}{l}{\small Av.Post.SD} & \multicolumn{1}{l}{\small Av.TS.SE} & \multicolumn{1}{l}{\small P.O.C} \\
		\hline
		1     & 2.00227 & 0.02015 & 0.00285 & 1 & 0.99973 & 0.01043 & 0.00012 & 1 \\
		0.98  & 1.99816 & 0.03117 & 0.00257 & 0.97 & 0.99815 & 0.01241 & 0.00049 & 0.98 \\
		0.95  & 1.99939 & 0.03827 & 0.00388 & 0.98 & 0.99921 & 0.00853 & 0.00029 & 0.99 \\
		0.92  & 2.00163 & 0.03539 & 0.00331 & 0.98  & 0.99941 & 0.00750 & 0.00025 & 0.98 \\
		\hline
	\end{tabular}%
	\caption{Posterior summaries for $\delta_0$ and $\delta_1$ using traditional BETEL and RBETEL methods in the linear regression setting. The reported values are average posterior mean (Av.Post.Mean), average posterior standard deviation (Av.Post.SD), average time series standard error (Av.TS.SE) and proportion of coverage (P.O.C)of the designed values $\delta^*_0=2$ and $\delta^*_1=1$.}
	\label{tab:Simulation2}%
\end{sidewaystable}%

\section{Empirical example} \label{s:Emp}
We apply the RBETEL method to estimate the relationship between log average brain weight (y) and log average body weight (x) for a dataset relating to sixty-five species of land animals. The same example is employed in the work of \cite{RV-1990unmasking}, who develop a Frequentist approach to unmask multivariate outliers and leverage points. Following \cite{RV-1990unmasking}, a linear regression model is employed to estimate the relationship, i.e.
\begin{equation}\label{eq:Emp_model}
	y_i = \delta_0 + \delta_1x_i + \epsilon_i \mbox{ for } i = 1,\dots,65,
\end{equation}
where the errors, $\epsilon_i$, are assumed to be iid.\\ 

\begin{figure}[H]
	\centering
	\includegraphics[width=0.6\textwidth]{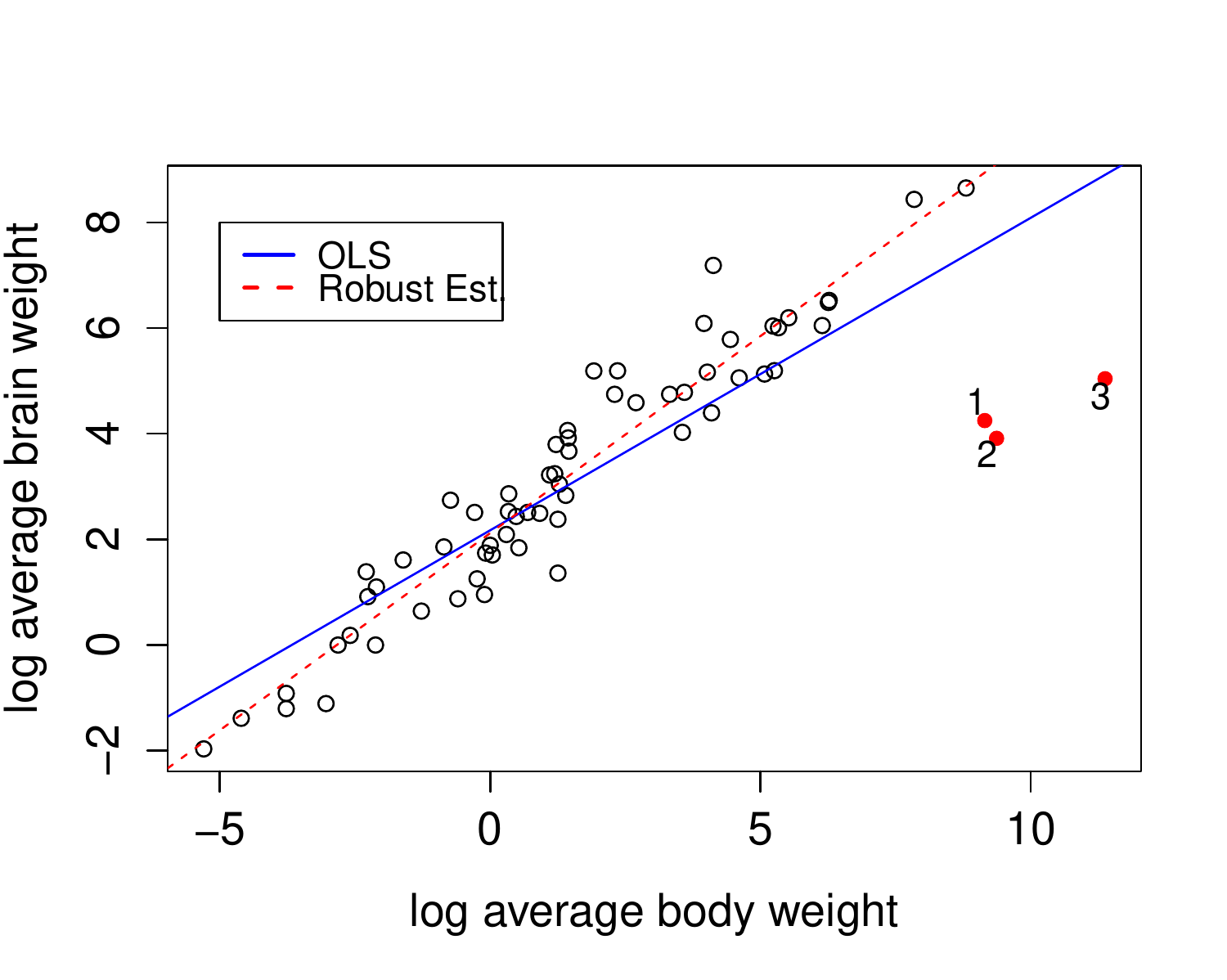}
	\caption{Plot of log average brain weights against log average body weights of sixty-five land animal species. The solid blue line is estimated by OLS. The dashed red line is obtained from a Frequentist robust M-estimation method.}
	\label{fig:Emp_dataplot}
\end{figure}

A scatter plot of the data is given in Figure \ref{fig:Emp_dataplot}, showing a clear linear relationship between the two variables for most of the observations. We use OLS estimation and robust M-estimation methods to produce a preliminary analysis of the linear relationship and estimation results are summarized in Table \ref{tab:RBETEL_Emp_FreqAna}. There are three obvious outliers at the right hand side of the plot and these outliers are leverage points. A regression line produced by robust M-estimation, with parameter estimates given by $\widehat{\delta}^{M}_0=2.12$ and $\widehat{\delta}^{M}_1=0.75$, seems to fit the majority of the data well. On the other hand, non-robust OLS estimation produces a regression line with an apparent lower slope due to the effect of the outliers, and the parameter estimates in this case are given by $\widehat{\delta}^{OLS}_0=2.17$ and $\widehat{\delta}^{OLS}_1=0.59$.

\begin{table}[H]
	\centering
	\begin{tabular}{l|rrrl}
		\hline
		\multicolumn{1}{r}{} &       & \multicolumn{1}{l}{OLS} &       &  \\
		\hline
		Parameter & \multicolumn{1}{l}{Estimate} & \multicolumn{1}{l}{Std.Error} & \multicolumn{1}{l}{t-statistic} & p-value \\
		\hline
		$\delta_0$ & 2.1717 & 0.1620 & 13.40 & $<2e-16$ \\
		$\delta_1$ & 0.5915 & 0.0411 & 14.37 & $<2e-16$ \\
		\hline
		\multicolumn{1}{r}{} &       &       &       &  \\
		\hline
		\multicolumn{1}{r}{} &       & \multicolumn{2}{l}{Robust estimation} &    \\
		\hline
		Parameter & \multicolumn{1}{l}{Estimate} & \multicolumn{1}{l}{Std.Error} & \multicolumn{1}{l}{t-statistic} & p-value \\
		\hline
		$\delta_0$ & 2.1175 & 0.0981 & 21.59 & $<2e-16$ \\
		$\delta_1$ & 0.7460 & 0.0249 & 29.94 & $<2e-16$ \\
		\hline
	\end{tabular}%
	\caption{Estimation results produced by OLS method and robust M-estimation method.}
	\label{tab:RBETEL_Emp_FreqAna}
\end{table}%

We employ the same moment conditions used in the simulation experiment discussed in Section \ref{s:Sim_Linear} for standard BETEL and RBETEL methods. For comparison, we also infer the linear relationship using a parametric Bayesian method based on model (\ref{eq:Emp_model}), and assuming a Student-t distribution for the errors with degree of freedom $\nu$, i.e. $\epsilon_i\overset{iid}{\sim}St(\nu)$. We assign a gamma prior to the degree of freedom parameter, $\nu$, with shape equal to $10$ and rate equal to $5$. We give a flat Normal prior with mean zero and variance equal to $100$ to the parameters $\delta_0$ and $\delta_1$. For RBETEL estimation, the prior distribution of the indicator vector, $s$, conditional on the probability parameter, $v$, is shown in Section 4.3.4.1. The prior of the probability parameter, $v$, given by (\ref{eq:Prior_v}) is a truncated Beta distribution and we set the hyperparameters to be $\alpha_0=30$ and $\beta_0=15$. We produce $50,000$ MCMC draws from the posterior distribution, and use the first $20,000$ draws to warm-up the chains and the subsequent $30,000$ draws to make inference.\\

\begin{figure}[H]
	\centering
	\begin{subfigure}[b]{0.45\textwidth}
		\includegraphics[width=\textwidth]{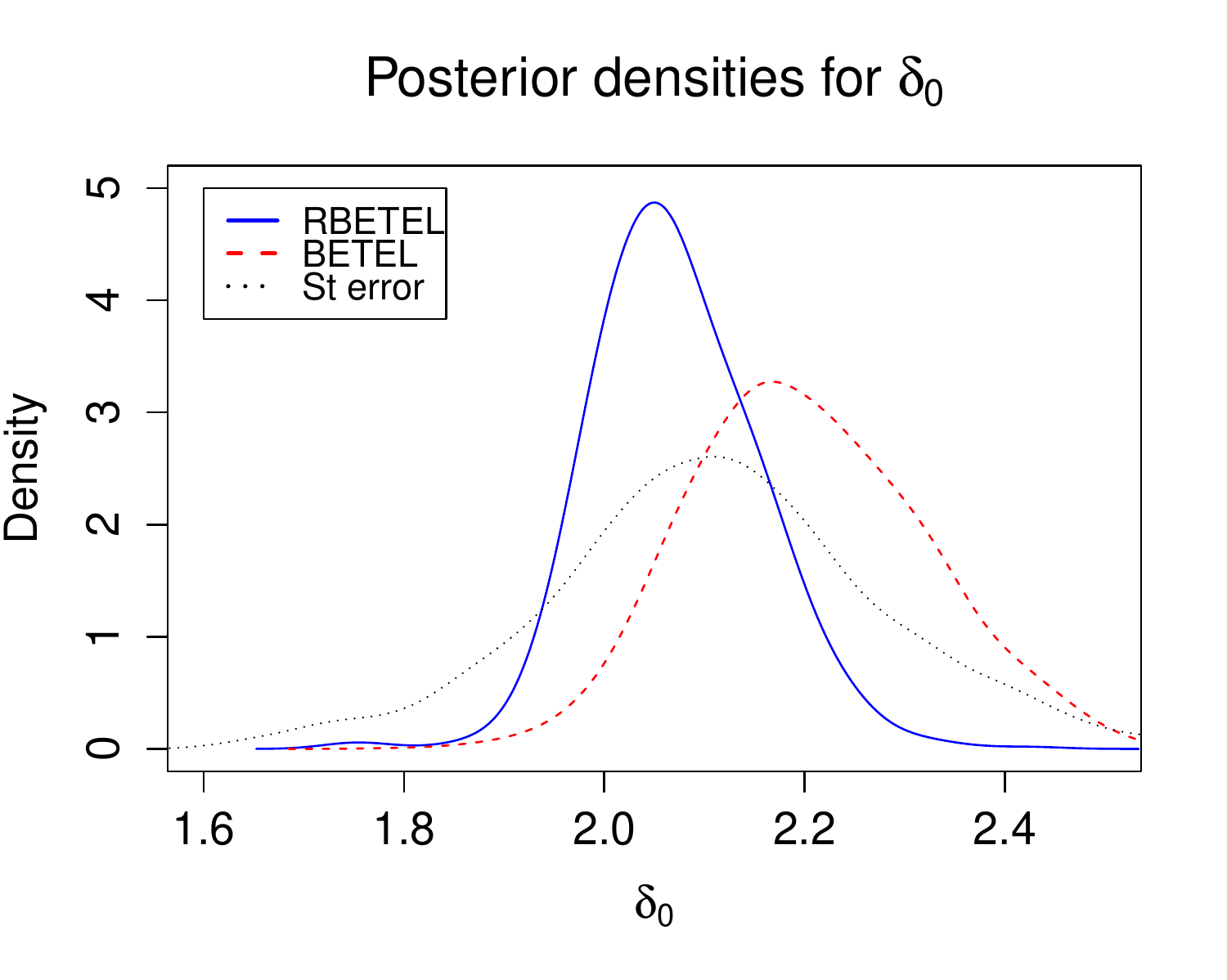}
		\label{fig:Emp_d0}
	\end{subfigure}
	\begin{subfigure}[b]{0.45\textwidth}
		\includegraphics[width=\textwidth]{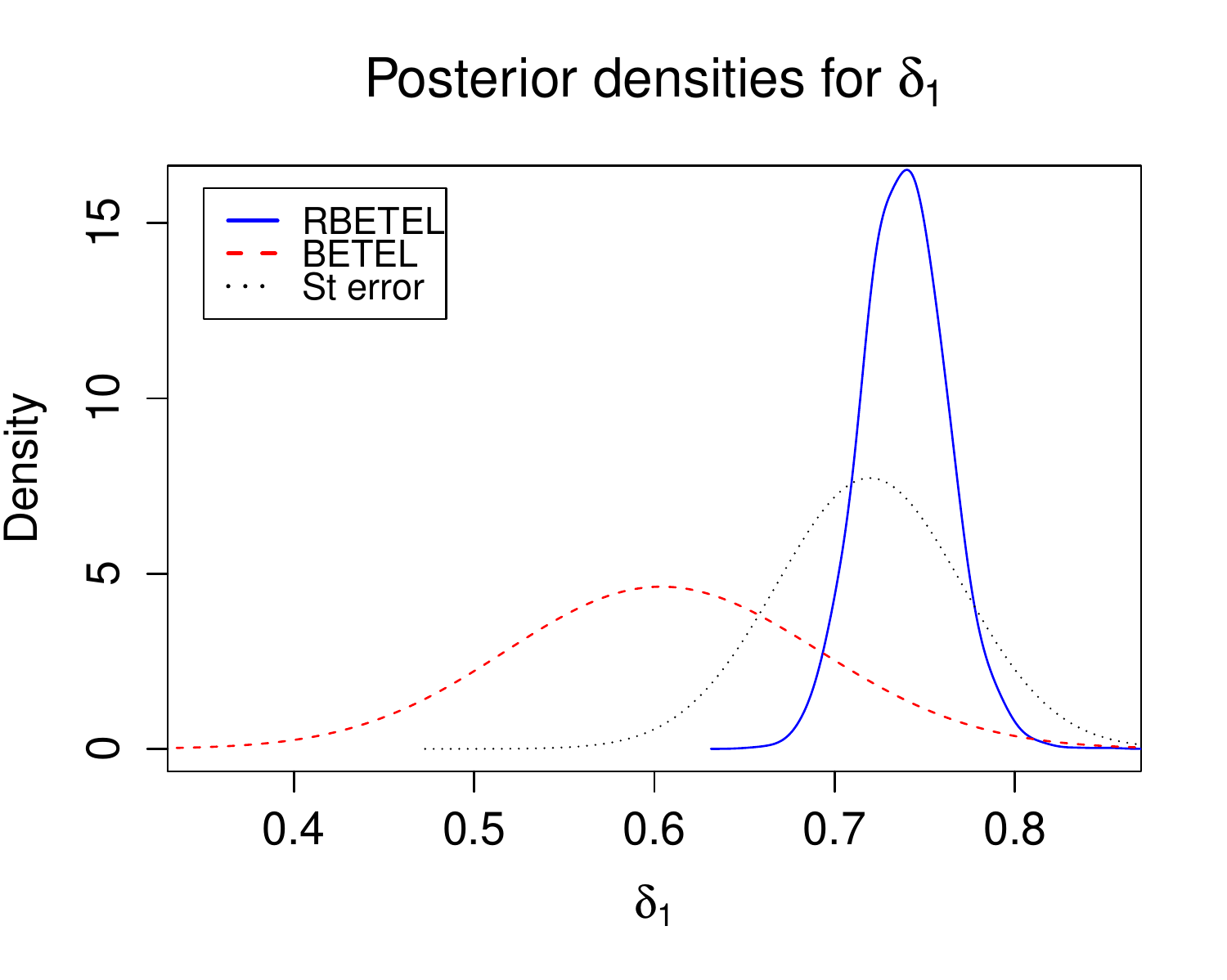}
		\label{fig:Emp_d1}
	\end{subfigure}
	\caption{Plot of posterior densities obtained using the BETEL method, RBETEL method and a traditional Bayesian parametric regression method under an assumption of Student-t errors.}
	\label{fig:Emp_posterior}
\end{figure}

The inferential results are summarized in Table \ref{tab:Emp_summary}, and Figure \ref{fig:Emp_posterior} provides the plots of the marginal posterior densities for the parameters $\delta_0$ and $\delta_1$, respectively. We can see that the RBETEL posterior mean estimates for the parameters, $\widehat{\delta}^{\textsl{\tiny{RBETEL}}}_0=2.1482$ and $\widehat{\delta}^{\textsl{\tiny{RBETEL}}}_1=0.7512$, are similar to the robust Frequentist estimates produced by M estimation. The posterior mean estimates produced by the standard BETEL method are given by $\widehat{\delta}^{\textsl{\tiny{BETEL}}}_0=2.2016$ and $\widehat{\delta}^{\textsl{\tiny{BETEL}}}_1=0.6055$, which are similar to those estimated by the non-robust OLS method. We find that the $95\%$ credible intervals for $\delta_1$ produced by the two methods considered here have little overlap. This suggests that inference about the relationship between the variables estimated by these two methods are significantly different. The parametric Bayesian approach assuming a Student-t distribution for the error term shows some extent of robustness with respect to outliers. The posterior mean estimates produced by this parametric approach for $\delta_0$ and $\delta_1$ are $2.1114$ and $0.7191$, respectively, and they are both between the corresponding estimates produced by the standard BETEL and RBETEL methods. Notably, however, the RBETEL posteriors here are much sharper than are the posteriors that result from the parametric and the BETEL methodologies.\\ 

\begin{figure}[H]
	\centering
	\includegraphics[width=0.6\textwidth]{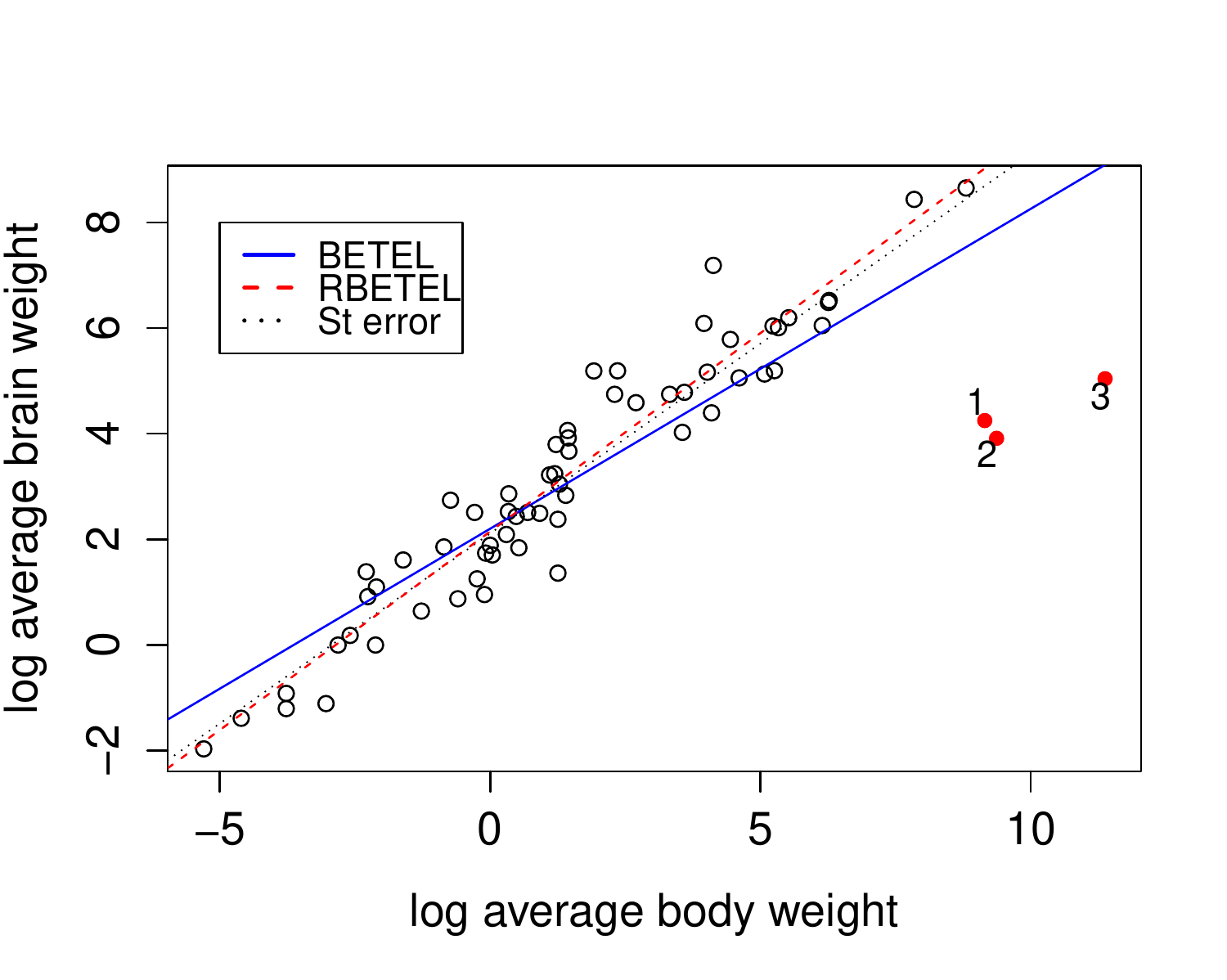}
	\caption{Regression lines obtained by the BETEL, RBETEL methods and traditional Bayesian parametric approach with Student-t errors. The red points $1$, $2$ and $3$ are outliers, as identified by the RBETEL method.}
	\label{fig:Emp_result_analysis}
\end{figure}

The estimated regression lines with parameter values given by the posterior means are added to the data plot in Figure \ref{fig:Emp_result_analysis}. We can see that the regression line estimated by the RBETEL method fits the majority of the observations well. The red points marked by number $1$, $2$ and $3$ in Figure \ref{fig:Emp_result_analysis}, are the outliers identified by the RBETEL method, while the corresponding estimated probabilities that these points are `good', i.e. $Pr(s_i=1)$, are $0.053$, $0.032$ and $0.001$, respectively. All other observations have estimated probabilities that they are `good' that are greater than $0.8$. Compared to the robust estimate produced by the RBETEL method, the regression line produced by the traditional BETEL method has an obviously lower slope due to the effect of the leverage points. The estimated regression line produced by the parametric Bayesian approach with Student-t errors is quite similar to the one produced by RBETEL method, but it has a lower slope. Presumably, the influence of outliers has been reduced by using the fat-tailed Student-t distribution, but it is unable to capture the asymmetry of the data distribution due to the presence of leverage points.\\ 

\begin{table}[htbp]
	\centering
	\begin{tabular}{l|rrrl}
		\hline
		\multicolumn{1}{r}{} &       & BETEL &  &  \\
		\hline
		Parameter & \multicolumn{1}{l}{Post.Mean} & $95\%$ C.I. & \multicolumn{1}{l}{Post.SD} & \multicolumn{1}{l}{T.S.S.E} \\
		\hline
		$\delta_0$ & 2.2016 & (1.9893, 2.4433) & 0.1171 & 0.0120 \\
		$\delta_1$ & 0.6055 & (0.5137, 0.7041) & 0.0496 & 0.0057 \\
		\hline
		\multicolumn{1}{r}{} &       &       &       &  \\
		\hline
		\multicolumn{1}{r}{} &       & RBETEL &       &  \\
		\hline
		Parameter & \multicolumn{1}{l}{Post.Mean} & $95\%$ C.I. & \multicolumn{1}{l}{Post.SD} & \multicolumn{1}{l}{T.S.S.E} \\
		\hline
		$\delta_0$ & 2.1482 & (1.9389, 2.3606) & 0.1098 & 0.0081 \\
		$\delta_1$ & 0.7512 & (0.7029, 0.8033) & 0.0256 & 0.0009\\
		\hline
		\multicolumn{1}{r}{} &       &       &       &  \\
		\hline
		\multicolumn{1}{r}{} &  & Parametric Bayesian & St error &  \\
		\hline
		Parameter & \multicolumn{1}{l}{Post.Mean} & $95\%$ C.I. & \multicolumn{1}{l}{Post.SD} & \multicolumn{1}{l}{T.S.S.E} \\
		\hline
		$\delta_0$ & 2.1114 & (1.7590, 2.4521) & 0.1676 & 0.0223 \\
		$\delta_1$ & 0.7191 & (0.6360, 0.8014) & 0.0221 & 0.0022\\
		\hline
	\end{tabular}%
	\caption{Summary of posterior inferential results given by the BETEL method, RBETEL method and Bayesian parametric estimation with Student-t errors.}
	\label{tab:Emp_summary}%
\end{table}%

\section{Conclusion and future work}
This paper develops a robust Bayesian exponentially tilted empirical likelihood method which produces robust inference with respect to outliers for models based on moment conditions. The new RBETEL method is built upon the BETEL framework proposed by \cite{Schennach-2005}. We point out that the empirical likelihood function in the BETEL framework is closely related to the EL ratio which is a nonparametric statistic for testing the validity of the moment conditions. Therefore, we reinterpret the BETEL method under the framework proposed by \cite{BHW-2016}, where a loss function is constructed based on the EL ratio. The resulting posterior for the parameters can be shown as the representation of the subjective uncertainty in the minimizer of the expected loss. Also inspired by the work of \cite{CSS-2017}, we propose to reformulate the moment conditions in the RBETEL approach by introducing an indicator vector which separates the complete dataset into good data and outliers. The RBETEL loss function is constructed as the EL ratio evaluated using a subset of the observations given by the indicator vector. Careful construction of key conditions which ensure that the RBETEL moment conditions are valid for the good data but invalid when outliers are present, ensures that the RBETEL approach is able to identify outliers from the dataset, and therefore produce robust inference about the parameters of interest.\\ 

We conduct simulation experiments to assess the performance of the RBETEL method under simple location and linear regression settings. The RBETEL method produces accurate posterior distributions for the model parameters when outliers are present, in the sense that all marginal posteriors are centered over the relevant designed parameter values. This is not always true for the traditional Bayesian method, particularly when large outliers are present. In addition, when the data does not contain outliers, the RBETEL method seems to perform as well as the original BETEL method. We also illustrate the RBETEL method under an empirical setting where the relationship between log average brain weight and log average body weight of sixty-five land animal species is of interest. The empirical data contain outliers which act as leverage points. The RBETEL method is able to produce robust inference regarding the relationship between the two variables and identify influential outliers in the data.\\ 

This paper only considers the RBETEL method under simple linear settings, but we expect that this method can be applied to a wide range of problems. Of course, this method is built for moment condition models, and some well known moment based problems that we will consider in the future include instrumental variable regression and stochastic volatility settings.\\
\clearpage

\section*{Appendix: derivation of the RBETEL posterior with given loss function}
In this appendix, we derive the joint RBETEL posterior for the parameters $\theta$ and $s$ conditional on given proportion of outliers, $v$. Given the loss function (\ref{eq:RBETEL_loss_function}), we aim to find a representation of subjective uncertainty in the parameter set $\left(\theta, s\right)$ that minimizes the expected loss given by
\begin{equation*}
	E^{F}\left[\widetilde{l}\left(\theta,s;x_{1:n}\right)\right]
	= \int \widetilde{l}\left(\theta,s;x_{1:n}\right) dF,
\end{equation*}
where $F$ is the unknown distribution that generates the data, given by (\ref{eq:RBETEL_data_DGP}).\\ 

Following the discussion in Section {\ref{s:BHW}}, we want to find the probability measures $\widetilde{\pi}(\theta\mid s,x_{1:n})$ and  $\widetilde{\pi}(s\mid v,x_{1:n})$ which are the minimizers of
\begin{equation}\label{eq:RBETEL_decision}
	\sum_{s} \left[ \left(\int \int \widetilde{l}(\theta,s;x_{1:n}) dF\ 
	\pi(\theta\mid s,x_{1:n}) d\theta\right) \pi(s\mid v, x_{1:n}) \right].
\end{equation} 
This can be done by minimizing the finite sample version of (\ref{eq:RBETEL_decision}) recursively subject to probability measures $\pi(\theta\mid s,x_{1:n})$ and $\pi(s\mid v, x_{1:n})$. Note that the joint probability measure $\widetilde{\pi}$ over the pair $(\theta, s)$ denotes the target minimizer of the loss function (\ref{eq:RBETEL_decision}).\\ 

First, we define a cumulative loss function
\begin{equation}\label{eq:RBETEL_cumulative_loss1}
	\begin{split}
		& L_1 \left(\pi(\theta\mid s,x_{1:n}); \pi(\theta), x_{1:n}\right) =\\ 
		&\int \widetilde{l}(\theta,s;x_{1:n}) \pi(\theta\mid s,x_{1:n}) d\theta + 
		\int \pi(\theta\mid s,x_{1:n}) \log\left(\frac{\pi(\theta\mid s,x_{1:n})}{\pi(\theta)}\right) d\theta\\
		&= \int \pi(\theta\mid s,x_{1:n}) \log\left(\frac{\pi(\theta\mid s,x_{1:n})}{\exp\{-\widetilde{l}(\theta,s;x_{1:n})\}\pi(\theta)}\right) d\theta.
	\end{split}
\end{equation}
According to \cite{BHW-2016}, this loss function is the Bayesian finite sample version of
\begin{equation*}
	\int \int \widetilde{l}(\theta,s;x_{1:n}) dF\  \pi(\theta\mid s,x_{1:n}) d\theta.
\end{equation*}
The cumulative loss function (\ref{eq:RBETEL_cumulative_loss1}) takes the form of KL divergence, so it is straightforward to find that the minimizer of (\ref{eq:RBETEL_cumulative_loss1}) is given by
\begin{equation}\label{eq:RBETEL_posterior1}
	\begin{split}
		\widetilde{\pi}(\theta\mid s,x_{1:n}) 
		&= \underset{\pi(\theta\mid s,x_{1:n})}{\arg\min}L_1 \left(\pi(\theta\mid s,x_{1:n}); \pi(\theta), x_{1:n}\right)\\ 
		&= \frac{\exp\left\{-\widetilde{l}(\theta,s;x_{1:n})\right\}\pi(\theta)}{\int \exp\left\{-\widetilde{l}(\theta,s;x_{1:n})\right\}\pi(\theta) d\theta}.
	\end{split}
\end{equation}
Substituting (\ref{eq:RBETEL_posterior1}) into (\ref{eq:RBETEL_cumulative_loss1}), we have the expression for the minimum of $L_1$ which is given by
\begin{equation}\label{eq:RBETEL_min_L1}
	\begin{split}
		\min L_1 \left(\pi(\theta\mid s,x_{1:n}); \pi(\theta), x_{1:n}\right)
		&= L_1 \left(\widetilde{\pi}(\theta\mid s,x_{1:n}); \pi(\theta), x_{1:n}\right)\\ 
		&= -\log \int \exp\left\{-\widetilde{l}(\theta,s;x_{1:n})\right\}\pi(\theta) d\theta.
	\end{split}
\end{equation}

Then the second cumulative loss function is corresponding to the expected loss
\begin{equation*}
	\sum_{s} \left[ \min \left(\int \int \widetilde{l}(\theta,s;x_{1:n}) \pi(\theta\mid s,x_{1:n}) d\theta\ dF\right) \pi(s\mid v, x_{1:n}) \right],
\end{equation*}
it is given by
\begin{equation}\label{eq:RBETEL_cumulative_loss2}
	\begin{split}
		& L_2 \left(\pi(s\mid v, x_{1:n}); \pi(s\mid v), x_{1:n}\right) =\\ 
		&\sum_{s}\pi(s\mid v, x_{1:n})
		\min L_1\left(\pi(\theta\mid s,x_{1:n}); \pi(\theta), x_{1:n}\right) + \sum_{s}\pi(s\mid v, x_{1:n}) \log\left(\frac{\pi(s\mid v, x_{1:n})}{\pi(s\mid v)}\right)\\
		&= \sum_{s} \pi(s\mid v, x_{1:n})\log\left(\frac{\pi(s\mid v, x_{1:n})}{\exp\left\{-\min L_1\left(\pi(\theta\mid s,x_{1:n}); \pi(\theta), x_{1:n}\right)\right\}\pi(s\mid v)}\right)\\
		&= \sum_{s} \pi(s\mid v, x_{1:n})\log\left(\frac{\pi(s\mid v, x_{1:n})}{\int \exp\left\{-\widetilde{l}(\theta,s;x_{1:n}) \pi(\theta) \right\}d\theta \mbox{ } \pi(s\mid v)}\right).
	\end{split}
\end{equation}

It is straightforward to find that the minimizer of (\ref{eq:RBETEL_cumulative_loss2}) is given by
\begin{equation}\label{eq:RBETEL_posterior2}
	\begin{split}
		\widetilde{\pi}(s\mid v,x_{1:n})
		&= \underset{\pi(s\mid v,x_{1:n})}{\arg\min}L_2 \left(\pi(s\mid v, x_{1:n}); \pi(s\mid v), x_{1:n}\right)\\ 
		&= \frac{\int \exp\left\{-\widetilde{l}(\theta,s;x_{1:n})\right\}\pi(\theta) d\theta\ \pi(s\mid v)}{\sum_{s}\int \exp\left\{-\widetilde{l}(\theta,s;x_{1:n})\right\}\pi(\theta) d\theta\ \pi(s\mid v)},
	\end{split}
\end{equation}
and then the minimum of $L_2$ is given by
\begin{equation}\label{eq:RBETEL_min_L2}
	\begin{split}
		\min L_2 \left(\pi(s\mid v, x_{1:n}); \pi(s\mid v), x_{1:n}\right)
		&= L_2 \left(\widetilde{\pi}(s\mid v, x_{1:n}); \pi(s\mid v), x_{1:n}\right)\\ 
		&= -\log \sum_{s}\int \exp\left\{-\widetilde{l}(\theta,s;x_{1:n})\right\}\pi(\theta) d\theta\ \pi(s\mid v).
	\end{split}
\end{equation}

Finally, the desired joint RBETEL posterior is given by the product of (\ref{eq:RBETEL_posterior1}) and (\ref{eq:RBETEL_posterior2}), i.e.
\begin{equation}\label{eq:RBETEL_posterior_complete}
	\begin{split}
		\pi_{\text{\tiny{RBETEL}}}(\theta,s\mid v,x_{1:n}) 
		&= \widetilde{\pi}(\theta\mid s,x_{1:n}) \widetilde{\pi}(s\mid v,x_{1:n})\\ 
		& \propto \pi(\theta)\pi(s\mid v)\exp\left\{-\widetilde{l}(\theta,s;x_{1:n})\right\}.
	\end{split}
\end{equation}

\clearpage

\bibliography{RBETEL}


\end{document}